\documentclass[10pt,a4paper]{article}

\usepackage{fullpage,enumerate}
\usepackage{amssymb,amsmath,mathrsfs}
\usepackage{xcolor}

\newtheorem{definition}{Definition}[section]
\newtheorem{lemma}[definition]{Lemma}
\newtheorem{proposition}[definition]{Proposition}
\newtheorem{theorem}[definition]{Theorem}

\newtheorem{remark}[definition]{Remark}

\newtheorem{example}[definition]{Example}
\newenvironment{proof*}{\smallskip\par\noindent\emph{Proof. }
 \ignorespaces}{\hfill$\Box$\smallskip\par\ignorespaces}
\newenvironment{proofsketch*}{\smallskip\par\noindent
 \emph{Sketch of proof. }\ignorespaces}
 {\hfill$\oslash$\smallskip\par\ignorespaces}

\newcommand{\C}{\ensuremath{\mathbb{C}}}
\newcommand{\N}{\ensuremath{\mathbb{N}}}
\newcommand{\R}{\ensuremath{\mathbb{R}}}
\newcommand{\Z}{\ensuremath{\mathbb{Z}}}

\newcommand{\A}{\ensuremath{\mathcal{A}}}

\title{\textbf{On separable states in relativistic quantum field theory}}
\author{Ko Sanders\thanks{E-mail:
jacobus.sanders@fau.de}}
\date{Department Mathematik, FAU
Erlangen-N\"urnberg, Cauerstra{\ss}e 11, 91058 Erlangen\\[2ex]
October 13, 2023}

\begin{document}
\maketitle

\section*{Abstract}

We initiate an investigation into separable, but physically reasonable, states in relativistic quantum field theory. In particular we will consider the minimum amount of energy density needed to ensure the existence of separable states between given spacelike separated regions. This is a first step towards improving our understanding of the balance between entanglement entropy and energy (density), which is of great physical interest in its own right and also in the context of black hole thermodynamics. We will focus concretely on a linear scalar quantum field in a topologically trivial, four-dimensional globally hyperbolic spacetime. For rather general spacelike separated regions $A$ and $B$ we prove the existence of a separable quasi-free Hadamard state.
% In particular such states do not have the Reeh-Schlieder property.
In Minkowski spacetime we provide a tighter construction for massive free scalar fields: given any $R>0$ we construct a quasi-free Hadamard state which is stationary, homogeneous, spatially isotropic and separable between any two regions in an inertial time slice $t=\mathrm{const.}$ all of whose points have a distance $>R$. We also show that the normal ordered energy density of these states can be made $\le 10^{31}\frac{m^4}{(mR)^8}e^{-\frac14mR}$ (in Planck units). To achieve these results we use a rather explicit construction of test-functions $f$ of positive type for which we can get sufficient control on lower bounds on $\hat{f}$.

\section{Introduction}

In relativistic quantum field theory (QFT), entanglement is ubiquitous. Many states are entangled between any two open, spacelike related regions of spacetime. Intuitively, creating entanglement costs no effort: one typically only needs to wait for the relativistic dynamics to create entanglement between given regions in space (at least in a static spacetime, where one can identify regions of space in the course of time without ambiguity). More mathematically, the entanglement can be seen as a consequence of the Reeh-Schlieder property, which was first discovered for the Wightman vacuum in Minkowski spacetime \cite{RS1961}, but also holds for many other states \cite{DM1971}, including such physically relevant states as thermal (KMS) states \cite{Str2000}, states of bounded energy \cite{Bor1968} and even for suitable states and spacetime regions in general globally hyperbolic curved spacetimes \cite{San2009}. We refer to the nice review by Witten \cite{Wit2018} and to \cite{Haag} for more information on the Reeh-Schlieder theorem and to Corollary 1 in Section 5.1. of \cite{HS2018} for a concrete proof that this implies the presence of entanglement.

Creating a state which is separable between given, spacelike separated regions of spacetime seems to be a less trivial task in relativistic QFT. Although it is known that one can construct such states mathematically, little seems to be known about the physical properties they may have. E.g., for massive free scalar fields, it is known that there are seperable states which are normal \cite{Buc1974}, which is a necessary condition to make them physically viable. This result should extend to quite general QFTs, due to the split property, cf.~\cite{BDF1987}. However, we are not aware of any results showing that such separable states can be e.g. Hadamard states with a finite energy density. It is the purpose of this paper to settle this question in the positive for massive free scalar fields in a rather broad setting. Moreover, in the case of Minkowski spacetime we show that such states may enjoy a large amount of symmetry and we also provide an upper bound for the energy density needed to have separable states. Note that the Hadamard property ensures that the states that we construct are also locally normal \cite{Ver1994}.

The entanglement entropy and energy density are two of the thermodynamic quantities that play a central role in black hole thermodynamics. A quantitative investigation into the balance between these quantities is of great interest to clarify the deep relations between thermodynamics, quantum theory and general relativity that black hole thermodynamics hints at. Furthermore, a better understanding of this balance could also lead to new insights that are relevant in quantum experiments, e.g. regarding the amount of energy needed to perform certain experiments. The current paper about separable states and their energy density can be seen as a small first step in these directions.

At a mathematical level, our analysis makes essential use of test-functions of positive type. A distribution $u$ on $\R^d$, $d\in\N$, is said to be of positive type iff $u(\bar{f}*\tilde{f})\ge0$ for all $f\in C_0^{\infty}(\R^d)$, where $*$ denotes the convolution, $\tilde{f}(x)=f(-x)$ and $C_0^{\infty}(\R^d)$ is the space of compactly supported smooth test-functions in $\R^d$.
% Note: $\int u(x)\bar{f}(x-y)\tilde{f}(y) dx dy = \int u(x)\bar{f}(x+y)f(y) dx dy = \int u(x-y)\bar{f}(x)f(y) dx dy$.
For this it is necessary and sufficient that $u$ is tempered and its Fourier transform is a positive measure, $\hat{u}\ge0$, of at most polynomial growth. This is the content of the Bochner-Schwartz theorem (cf.~\cite{RS1980} Thm.~IX.10). Distributions of positive type are useful in various applications, because they define a (semi-definite) inner product on $C_0^{\infty}(\R^d)$.

It is well-known that one can construct test-functions $f\in C_0^{\infty}(\R^d)$ of positive type and with support contained in an arbitrarily small neighbourhood of the origin. Indeed, if $g\in C_0^{\infty}(\R^d)$ is an even, real-valued function, then $\hat{g}$ is even, real-valued and real analytic. Consequently, $f:=g*g$ is of positive type with $\hat{f}=\hat{g}^2\ge0$ real analytic. 

Much is also known about the upper bounds on $\hat{f}$ for $f\in C_0^{\infty}(\R^d)$ of positive type. E.g., in the construction above, if
$|\hat{g}(k)|\le C e^{-|k|\epsilon(|k|)}$ for all $k\in\R^d$ and some $C>0$ and $\epsilon:\R_{\ge0}\to\R$, then $|\hat{f}(k)|\le C^2 e^{-2|k|\epsilon(|k|)}$. Using this one can show that the positive type condition can be imposed essentially independently of any requirements on the fall-off behaviour of $\hat{f}$.

In this paper we will need to focus instead on lower bounds on $\hat{f}$. It is easy to show that one can make $\hat{f}$ strictly positive: returning to the construction above, if $g$ is not identically zero, then $\hat{f}>0$ almost everywhere, so $\hat{h}=\hat{f}*\hat{f}$ is even and $\hat{h}>0$ everywhere. Note that $h=\frac{1}{(2\pi)^d}(g*g)^2$ is still smooth and we may choose its support as small as we like by shrinking the support of $g$.

Some information about lower bounds on $\hat{f}$ can be inferred from \cite{FF2015}, who constructed test functions of positive type on the real line with arbitrarily small support and bounds on their asymptotic behaviour: for any $\alpha\in(0,1)$ and $\beta,\delta>0$ they found a real-valued $f\in C_0^{\infty}(\R)$ with support in $[-2\delta,2\delta]$ such that $\hat{f}(k)=e^{-2\beta|k|^{\alpha}}\left(C+O(|k|^{\alpha-1})\right)$ for $k\not=0$ and some constant $C$
(cf.~Eqn.~(35) loc.cit.). In applications this estimate may not always be precise enough, however, as we will see in Section \ref{sec:QFT}.

In Section \ref{sec:positivity} we will construct test-functions of positive type with arbitrarily small support near the origin and satisfying lower bounds of a quite general form that hold for all $k\in\R^d$. We then apply some of our constructions in Section \ref{sec:QFT} to construct nice quantum states in relativistic QFT with surprisingly little entanglement. In Minkowski spacetime we also estimate the energy density of the states that we construct. (For comparison, the results of \cite{FF2015} would only allow the conclusion that the energy density of our states is finite.) As a matter of notation, we sometimes write the shorthand $\int f$ to denote integration of $f$ over an entire space with a naturally associated volume form (e.g. $\int_{\R^d}f(x)\mathrm{d}^dx$).

\section{Test functions of positive type}\label{sec:positivity}

In this section we will develop some results on rotation invariant distributions and test functions of positive type in the Euclidean space $\R^d$, $d\in\N$. By rotation invariance we will mean invariance under $O(d)$, so if $d=1$ the distribution should be even. Our aim is to control the lower bounds of the Fourier transform $\hat{f}(k)=\int e^{-ik\cdot x}f(x)\mathrm{d}x$ of a test-function $f$ of positive type. At the same time we will ensure that $f$ is pointwise positive and rotation invariant. Note that a simultaneous control of local properties of $f$ and of its Fourier transform $\hat{f}$ can in general be challenging, as the following lemma illustrates.
\begin{lemma}\label{lem:negativity}
Let $f\in C_0^{\infty}(\R^d)$ be of positive type. Then $f(0)\ge0$ with equality iff $f\equiv0$. For any Cartesian coordinate $x^1$, 
$\partial_{x^1}^2f(0)\le 0$ with equality iff $f\equiv 0$.
\end{lemma}
\begin{proof*}
In terms of the Fourier transform $\hat{f}$ of $f$ we have $f(0)=\frac{1}{(2\pi)^d}\int \hat{f}(k)\mathrm{d}^dk\ge0$ because $\hat{f}(k)\ge0$. We have equality iff $\hat{f}\equiv0$ which would give $f\equiv0$. Similarly, $\partial_{x_1}^2f(0)=\frac{1}{(2\pi)^d}\int -k_1^2\hat{f}(k)\mathrm{d}^dk\le 0$ and equality holds iff $k_1^2\hat{f}(k)\equiv 0$. Because $\hat{f}$ is real analytic this yields $\hat{f}(k)\equiv 0$ and hence $f\equiv 0$.
\end{proof*}

\subsection{A construction of test functions of positive type in one dimension}\label{ssec:1dconstruction}

We first consider $d=1$ and we apply the procedure from the Introduction to a multiple of the characteristic function $\chi$ of the interval $\left[-\frac12,\frac12\right]$. This leads to the function
\begin{align}\label{def:eta}
\eta(x)&:=\frac32(\chi*\chi)^2(x)=
\begin{cases}
\frac32(1-|x|)^2&\mathrm{if}\ |x|\le 1\\
0&\mathrm{if}\ |x|> 1
\end{cases}
\end{align}
with Fourier transform
\begin{align}\label{eqn:etahat}
\hat{\eta}(k)&
% =\frac32\int_{-1}^1e^{-ixk}(1-|x|)^2\mathrm{d}x
=3\int_0^1 \cos(xk)(1-x)^2\mathrm{d}x
% \notag\\
% &
=\begin{cases}
\frac{6}{k^3}(k-\sin(k))&\mathrm{if}\ k\not=0\\
1&\mathrm{if}\ k=0
\end{cases}\,,
\end{align}
where the normalisation is chosen such that $\int\eta=1$. We will use $\hat{\eta}$ as a basic building block in the constructions that follow, so it is worthwhile to establish some further properties of this function.

\begin{lemma}\label{lem:etaprop}
The function $\eta$ defined in (\ref{def:eta}) satisfies
\begin{align}
\frac{1}{1+\beta k^2}&\le \hat{\eta}(k)\le \frac{1}{1+\alpha k^2}\,,\notag
\end{align}
for all $k\in\R$, where $\alpha=\frac{1}{20}$ and $\beta=\frac{7}{40}$.
\end{lemma}

\begin{proof*}
Because $\hat{\eta}(k)$ is even it suffices to prove the estimates for $k\ge0$. On $k>0$ we have $k>\sin(k)$, so from (\ref{eqn:etahat}) we see that $\hat{\eta}(k)>0$ if $k\ge0$. Integrating $k-\sin(k)>0$ twice from $0$ to $k>0$ we also find
% $\frac12k^2+\cos(k)-1>0$ and
$\frac16k^3+\sin(k)-k>0$, which can be rearranged to show that $\hat{\eta}(k)<1$ on $k>0$. It follows that $f(k):=k^{-2}\left(\hat{\eta}^{-1}(k)-1\right)$ 
is smooth and even and $f(k)>0$ for $k\not=0$. Because $\hat{\eta}(0)=1$ we have
\begin{align}
\lim_{k\to0}f(k)&=\lim_{k\to0}\frac{1}{k^2\hat{\eta}(k)}\left(1-\frac{6(k-\sin(k))}{k^3}\right)
=\lim_{k\to0}6\frac{\frac16k^3-k+\sin(k)}{k^5}=6\frac{1}{5!}=\frac{1}{20}\notag
\end{align}
and
\begin{align}
\lim_{k\to\pm\infty}f(k)&=\lim_{k\to\pm\infty}k^{-2}\left(\frac{k^3}{6(k-\sin(k))}-1\right)
=\lim_{k\to\pm\infty}\frac{1}{6(1-k^{-1}\sin(k))}=\frac16\,,\notag
\end{align}
so $\displaystyle{\alpha':=\inf_kf(k)>0}$ and $\displaystyle{\beta':=\sup_kf(k)<\infty}$. We then have $1+\alpha' k^2\le\hat{\eta}^{-1}(k)\le 1+\beta' k^2$ for all $k$ and hence
$\frac{1}{1+\beta' k^2}\le\hat{\eta}(k)\le \frac{1}{1+\alpha' k^2}$.

We now show that $\alpha'=\alpha=\frac{1}{20}$. We clearly have $\displaystyle{\alpha'\le\lim_{k\to0}f(k)=\frac{1}{20}}$. The reverse inequality holds iff
$g(k):=k^2(k-\sin(k))\left(f(k)-\frac{1}{20}\right)$ is non-negative on $k>0$. Note that
\begin{align}
g(k)&=\sin(k)-k+\frac16k^3-\frac{1}{20}k^2(k-\sin(k))=\frac{7}{60}k^3-k+\left(1+\frac{1}{20}k^2\right)\sin(k)\,.\notag
\end{align}
Using $\sin(k)\ge-1$ we have $g(k)\ge \frac{7}{60}k^3-\frac{1}{20}k^2-k-1$ and using elementary methods one shows that $g(k)>0$ for $k\ge 4$.
% E.g., set $h(k):=\frac{7}{60}k^3-\frac{1}{20}k^2-k-1$ and note that $h(4)>0$, $h'(4)>0$, $h''(4)>0$ and $h'''(k)>0$.
On $k\ge0$ we can also integrate $-\cos(k)\ge -1$ seven times from $0$ to $k$ to get $\sin(k)\ge k-\frac{1}{3!}k^3+\frac{1}{5!}k^5-\frac{1}{7!}k^7$. Substituting this estimate in the definition of $g(k)$ gives after a little algebra $g(k)\ge \frac{k^7}{20\cdot 7!}(22-k^2)$. It follows that $g(k)\ge0$ when $k\in[0,\sqrt{22}]$. Putting both partial results together we have $g(k)\ge0$ on $[0,\sqrt{22}]\cup [4,\infty)=[0,\infty)$. Hence, $\alpha'=\frac{1}{20}$.

Finally we show that $\beta'\le \beta=\frac{7}{40}$, i.e. $f(k)\le\frac{7}{40}$ for all $k\in\R$, which is equivalent to $h(k)\ge0$ with
\begin{align}
h(k)&:=7k^2(k-\sin(k))-40\left(\sin(k)-k+\frac16k^3\right)\notag
\end{align}
for all $k\ge0$. We provide separate estimates for $k$ in the intervals $\left[0,\frac72\right]$, $[\pi,2\pi]$, $[2\pi,3\pi]$, $[3\pi,4\pi]$ and $[12,\infty)$, which together complete the proof. Note that $h(k)=\frac13k^3+40k-40\sin(k)-7k^2\sin(k)$ and using
$\sin(k)\le 1$ we find $h(k)\ge \frac13k^3-7k^2+40k-40=:p_1(k)$. The polynomial $p_1$ is strictly increasing for $k\ge 10$, because $p_1'(k)=k^2-14k+40=(k-4)(k-10)$. Since $p_1(12)=576-1008+480-40=8>0$ we find that $h(k)\ge p_1(k)>0$ for all $k\in[12,\infty)$. For $k\in[\pi,2\pi]\cup[3\pi,4\pi]$ we have $\sin(k)\le0$ and hence $h(k)\ge \frac13k^3+40k>0$ as required. For $k\ge 0$ we integrate the inequality $\cos(k)\le 1$ repeatedly from $0$ to $k$ to find $\sin(k)\le k-\frac{1}{3!}k^3+\frac{1}{5!}k^5$ and $\sin(k)\le k-\frac{1}{3!}k^3+\frac{1}{5!}k^5-\frac{1}{7!}k^7+\frac{1}{9!}k^9$. Hence,
\begin{align}
h(k)&=7k^2(k-\sin(k))-40\left(\sin(k)-k+\frac16k^3\right)\notag\\
&\ge 7k^2\left(\frac{1}{3!}k^3-\frac{1}{5!}k^5\right)-40\left(\frac{1}{5!}k^5-\frac{1}{7!}k^7+\frac{1}{9!}k^9\right)\notag\\
&=k^5p_2(k)\notag
\end{align}
where $p_2(k):=\frac56-\frac{254}{7!}k^2-\frac{40}{9!}k^4$. Now $p_2$ is decreasing on $k\ge0$ and
$p_2\left(\frac72\right)=\frac{7\cdot 20677}{2\cdot 9!}>0$, so $h(k)\ge k^5p_2(k)\ge0$ for $k\in \left[0,\frac72\right]$. Finally, for $k\in[2\pi,3\pi]$ we can integrate the inequality $\sin(k)\le 1$ from $\frac52\pi$ to $k$ repeatedly to find 
% $-\cos(k)\le k-\frac52\pi$
% $-\sin(k)\le -1+\frac12\left(k-\frac52\pi\right)^2$
% $\cos(k)\le -\left(k-\frac52\pi\right)+\frac16\left(k-\frac52\pi\right)^3$
$\sin(k)\le 1-\frac12\left(k-\frac52\pi\right)^2+\frac{1}{4!}\left(k-\frac52\pi\right)^4$ and hence, writing $l:=k-\frac52\pi$,
\begin{align}
h(k)&=\frac13k^3+40k-40\sin(k)-7k^2\sin(k)\notag\\
&\ge
\frac13\left(l+\frac52\pi\right)^3-7\left(l+\frac52\pi\right)^2+40\left(l+\frac52\pi\right)
-40\left(1-\frac12l^2+\frac{1}{4!}l^4\right)\notag\\
&>-\frac53l^4+\frac13l^3+20l^2+bl+3\notag\\
&=:p_3(l)\,,\notag
\end{align}
where $b=40-35\pi+\frac{25}{4}\pi^2$ and we estimated $\frac52\pi-7>0$ and $-40+100\pi-\frac{175}{4}\pi^2+\frac{125}{24}\pi^3>3$. When
$|l|\le\frac12\pi$ we can use the fact that $-\frac53l^2+\frac13l\ge-\frac53\left(\frac12\pi\right)^2-\frac13\cdot\frac12\pi>-5$ to see that
\begin{align}
p_3(l)&\ge -5l^2+20 l^2+bl+3=15l^2+bl+3=:p_4(l)\,.\notag
\end{align}
With an elementary computation one now shows that the discriminant $b^2-180$ of $p_4(l)$ is negative, which means that $p_4(l)>0$ for all $l\in\R$.
When $k\in[2\pi,3\pi]$ we then have $|l|\le\frac12\pi$ and $h(k)\ge p_3(l)\ge p_4(l)>0$.

We have now shown the desired estimate $h(k)\ge0$ for all $k\ge0$ and hence $\beta'\le\beta=\frac{7}{40}$. The inequality with $\beta$ then follows.
\end{proof*}

\begin{remark}
The value $\beta=\frac{7}{40}=0.175$ may not be sharp, but it will be good enough for our purposes. The exact value of $\beta'$ appearing in the proof of Lemma \ref{lem:etaprop} is harder to determine. It is clear from the proof that $\beta'\le\beta$ and $\beta'\ge\frac16$ and we also have
\begin{align}
\beta'&\ge f\left(\frac52\pi\right)
=\frac{48-120\pi+125\pi^3}{150\pi^2(5\pi-2)}
\simeq 0.174772\,.\notag
\end{align}
\end{remark}

To get smooth test functions of positive type we will consider repeated convolutions of functions of the form $\frac{1}{a_n}\eta\left(\frac{x}{a_n}\right)$ for suitable coefficients $a_n>0$ (cf.~Theorem 1.3.5 in \cite{Hormander} for an analogous construction of test functions that does not guarantee positive type). For the Fourier transforms we then find products of the functions $\hat{\eta}(a_nk)$. The proof of the following proposition is essentially standard, but is included for completeness.

\begin{proposition}\label{prop:defg}
Let $\{a_n\}_{n\in\N}$ be a sequence in $(0,\infty)$ such that the series $a=\sum_{n=1}^{\infty}a_n$ converges. For $n\in\N$ let
$\eta_n(x):=\frac{1}{a_n}\eta\left(\frac{x}{a_n}\right)$ and define $g_n(x)$ recursively by $g_{n+1}=g_n*\eta_{n+1}$ and $g_1=\eta_1$. Then $g_n$ converges uniformly to a smooth, real, even, non-negative function $g$ of positive type supported in $[-a,a]$ with $\int g=1$. Furthermore,
\begin{align}
\hat{g}(k)&=\prod_{n=1}^{\infty}\hat{\eta}(a_nk)\notag
\end{align}
and for all $l\in\N$ we have the following estimates on $L^1$ and suppremum norms:
\begin{align}
\|g\|_{\infty}&\le g(0)\le \frac{3}{2a_1}\notag\\
\|g\|_1&\le1\notag\\
\|g^{(2l)}\|_{\infty}&\le (-1)^lg^{(2l)}(0)\notag\\
\|g^{(l)}\|_1&\le\frac{3^l}{a_1\cdots a_l}\notag\\
\|g^{(l)}\|_{\infty}&\le \frac{1}{a_1}\frac{3^l}{a_1\cdots a_l}\,.\notag
\end{align}
\end{proposition}
\begin{proof*}
Because all $\eta_n$ are non-negative and even functions, so are the $g_n$. $\eta_n$ is supported in $[-a_n,a_n]$ and hence $g_n$ is supported in
$\left[-\sum_{j=1}^na_j,\sum_{j=1}^na_j\right]$. Moreover,
\begin{align}
\widehat{g_n}(k)&=\prod_{j=1}^n\widehat{\eta_j}(k)=\prod_{j=1}^n\hat{\eta}(a_jk)\le\prod_{j=1}^n\frac{1}{1+\alpha a_j^2k^2}\notag
\end{align}
by Lemma \ref{lem:etaprop}. Because $0\le\widehat{\eta_n}(k)\le 1$ it follows in particular that $g_n$ is of positive type, $0\le g_{n+1}(k)\le g_n(k)$ for all $n\in\N$ and $k\in\R$.

For any pair of bounded integrable functions $u,v$ we have $\|u*v\|_1\le \|u\|_1\|v\|_1$ and $\|u*v\|_{\infty}\le\|u\|_{\infty}\|v\|_1$. Because
$\|\eta_n\|_1=1$ and $\|\eta_n\|_{\infty}=\frac{3}{2a_n}$ it follows that $\|g_n\|_{\infty}\le \frac{3}{2a_1}$, whereas
$\|g_n\|_1=\hat{g}_n(0)=1$ because $g_n$ is non-negative.

To see that the functions $g_n$ converge uniformly we pick an arbitrary $\epsilon>0$. Because $\widehat{\eta_1}$ is integrable we can then find a $K>0$ such
that $\int_{|k|\ge K}\widehat{\eta_1}(k)\mathrm{d}k\le\frac{\epsilon}{4}$ and hence
\begin{align}
\left|\int_{|k|\ge K}\widehat{g_n}(k)-\widehat{g_m}(k)\mathrm{d}k\right|&\le\frac{\epsilon}{2}\notag
\end{align}
for all $m,n\in\N$. On the other hand, since $\hat{\eta}$ is smooth and even there is a $\delta>0$ such that $|k|<\delta$ implies $0\le 1-\hat{\eta}(k)\le\frac{\epsilon}{4K^2a}|k|$. Because $\sum_na_n$ converges, $\displaystyle{\lim_{n\to\infty}a_n}=0$, so there is an $N\in\N$ such that $n\ge N$ implies $a_nK<\delta$. When $m\ge n$ and $\epsilon\in(0,4)$ it then follows that
\begin{align}
\left|\int_{|k|<K}\widehat{g_n}(k)-\widehat{g_m}(k)\mathrm{d}k\right|&\le
\int_{|k|<K}|\widehat{g_n}(k)|\left(1-\prod_{j=n+1}^m\hat{\eta}(a_jk)\right)\mathrm{d}k\notag\\
&\le \int_{|k|<K}1-\prod_{j=n+1}^m\left(1-\frac{\epsilon}{4K^2a}a_j|k|\right)\mathrm{d}k\notag\\
&\le \int_{|k|<K}\sum_{j=n+1}^m\frac{\epsilon}{4K^2a}a_j|k|\mathrm{d}k\notag\\
&\le 2K\sum_{j=n+1}^m\frac{\epsilon}{4Ka}a_j<\frac{\epsilon}{2}\,,\notag
\end{align}
where we used the fact that $\prod_{j=1}^l(1-\epsilon_j)\ge1-\sum_{j=1}^l\epsilon_j$ for all $\epsilon_1,\ldots,\epsilon_l\in(0,1)$.
% Proof:
% By induction over $l\ge1$. For $l=1$ we trivially have $1-\epsion_1\ge1-\epsion_1$. When the inequality holds for some $l\in\N$ we have
% \begin{align}
% $\prod_{j=1}^{l+1}(1-\epsilon_j)&\ge \left(1-\sum_{j=1}^l\epsilon_j\right)(1-\epsilon_{l+1})\notag\\
% &=1-\sum_{j=1}^{l+1}\epsilon_j+\epsilon_{l+1}\sum_{j=1}^l\epsilon_j\ge1-\sum_{j=1}^{l+1}\epsilon_j\,.\notag
% \end{align}
Hence $\|g_n-g_m\|_{\infty}\le\frac{1}{2\pi}\|\widehat{g_n}-\widehat{g_m}\|_1<\epsilon$, so the functions $g_n$ form a Cauchy sequence w.r.t. the supremum norm and they converge uniformly to a limit $g$. It follows that $g$ is non-negative and even with support in $[-a,a]$, $\|g\|_{\infty}\le\frac{3}{2a_1}$. It also follows that $\|\widehat{g_n}-\widehat{g_m}\|_{\infty}\le 2a\|g_n-g_m\|_{\infty}$ converges uniformly to
$\hat{g}(k)=\prod_{n=1}^{\infty}\hat{\eta}(a_nk)$ and in particular $g$ is of positive type and $\|g\|_1=\hat{g}(0)=1$.

It remains to prove smoothness and the estimates on the derivatives of $g$. The weak derivative of $\eta(x)$ is given by
\begin{align}
\eta'(x)&=\begin{cases}
3(x+1)&\mathrm{if}\ -1\le x\le 0\\
3(x-1)&\mathrm{if}\ 0< x\le 1\\
0&\mathrm{if}\ |x|>1
\end{cases}\,,
\end{align}
(Note that the derivative at the point $x=0$ is not defined, but no multiples of $\delta$ distributions supported at this point appear.) It follows that $\eta_n'(x)=\frac{1}{a_n^2}\eta'\left(\frac{x}{a_n}\right)$ and we can compute
\begin{align}
\|\eta'_n\|_1=\frac{3}{a_n}\,,&\quad \|\eta_n\|_{\infty}=\frac{3}{a_n^2}\,.\notag
\end{align}
When $l\in\Z_{\ge0}$ and $n>l$ we have $g_n^{(l)}=\eta_1'*\cdots *\eta_l'*\eta_{l+1}*\cdots*\eta_n$ and by essentially the same argument as above
the functions $g_n^{(l)}$ converge uniformly to a function, which must be $g^{(l)}$. Furthermore, $\|g_n^{(l)}\|_1\le \frac{3^l}{a_1\cdots a_l}$ and $\|g_n^{(l)}\|_{\infty}\le \frac{1}{a_1}\frac{3^l}{a_1\cdots a_l}$, which yields $\|g^{(l)}\|_1\le\frac{3^l}{a_1\cdots a_l}$ and $\|g^{(l)}\|_{\infty}\le \frac{1}{a_1}\frac{3^l}{a_1\cdots a_l}$ in the limit.
% The weak second derivative is $\eta''(x)=-6\delta_0(x)+3\chi_[-1,1](x)$, where $\chi_[-1,1]$ is the characteristic function of the interval $[-1,1]$.
% Similarly, $\eta_n''(x)=\frac{1}{a_n^3}\eta''\left(\frac{x}{a_n}\right)=\frac{-6}{a_n^2}\delta_0(x)+\frac{3}{a_n^3}\chi_{[-a_n,a_n]}$.
% However, expressing $g^{(k)}$ in terms of $\eta_n''$ does not seem to yield better bounds.
\end{proof*}

\begin{example}\label{ex:Gevrey0}
For any $R>0$ and $\rho>1$ we can consider the sequence $\{a_n\}_{n\in\N}$ given by $a_n:=R\frac{\rho-1}{\rho}n^{-\rho}$. We then have
\begin{align}
a&:=\sum_{n=1}^{\infty}a_n=R\frac{\rho-1}{\rho}\sum_{n=1}^{\infty}n^{-\rho}
\le R\frac{\rho-1}{\rho}\left(1 + \int_1^{\infty}x^{-\rho}\mathrm{d}x\right)
=R\frac{\rho-1}{\rho}\left(1 + \frac{1}{\rho-1}\right)=R\,.\notag
\end{align}
The function $g$ defined in Proposition \ref{prop:defg} is therefore supported in the interval $[-R,R]$. Furthermore, for any $l\in\N$ we have
\begin{align}
\|g^{(l)}\|_1&\le \left(\frac{3\rho}{R(\rho-1)}\right)^ll!^{\rho}\,.\notag
\end{align}
This means that $g$ is in the Gevrey class of order $\rho>1$ and therefore
\begin{align}
|\hat{g}(k)|&\le c^{-1}e^{-c|k|^{\frac{1}{\rho}}}\notag
\end{align}
for some $c>0$ and all $k\in\R$, cf.~Appendix \ref{sec:Gevrey}.
% Equivalently, there is a $c>0$ such that $|\hat{g}(k)|\le c^{-1}\exp\left(-c|k|\cdot(1+|k|)^{\frac{1}{\rho}-1}\right)$ for all $k\in\R$.
% Proof:
% By continuity of $\hat{g}$ it suffices to consider $|k|\ge1$. There we have
% $|k|(1+|k|)^{\frac{1}{\rho}-1}\le |k|\cdot|k|^{\frac{1}{\rho}-1}=|k|^{\frac{1}{\rho}}\le 2^{1-\frac{1}\rho}}|k|(1+|k|)^{\frac{1}{\rho}-1}$
% because $\frac{1}{\rho}-1<0$.
\end{example}

In Section \ref{ssec:1dbounds} below we will investigate how well we can control the functions $\hat{g}(k)$ as constructed in Proposition \ref{prop:defg}. For this purpose we now establish a lemma.

\begin{lemma}\label{lem:defA}
Let $\{a_n\}_{n\in\N}$ be a sequence in $(0,\infty)$ such that the series $a=\sum_{n=1}^{\infty}a_n$ converges. Then the series
\begin{align}
A(k)&:=\sum_{n=1}^{\infty}\log(1+a_n^2k^2)\,,\label{eqn:defA}
\end{align}
converges uniformly on compact sets to a continuous function $A:\R\to\R$ with the following properties:
\begin{enumerate}[(i)]
\item $0\le A(k)\le a|k|$ for all $k\in\R$,
\item $A(k)$ is even and strictly increasing on $k\ge0$,
\item $\displaystyle{\lim_{k\to\infty}A(k)=\infty}$ and $\displaystyle{\lim_{k\to\infty}\frac{A(k)}{|k|}=\lim_{k\to0}\frac{A(k)}{|k|}=0}$,
\item $\int_0^{\infty}\frac{A(k)}{k^2}\mathrm{d}k=\pi a$,
\item for $\hat{g}(k)$ as defined in Proposition \ref{prop:defg} and for $\alpha=\frac{1}{20}$ and $\beta=\frac{7}{40}$ as in Lemma \ref{lem:etaprop},
\begin{align}
e^{-A(\sqrt{\beta}k)}&\le \hat{g}(k)\le e^{-A(\sqrt{\alpha}k)}\,.\notag
\end{align}
\end{enumerate}
\end{lemma}
\begin{proof*}
We consider the real Banach algebra $\mathcal{B}:=C_0^0((0,\infty),\R)$ of continuous functions on $(0,\infty)$ that vanish at $0$ and infinity, endowed with the supremum norm $\|.\|_{\infty}$. The function $f(r):=\frac{1}{r}\log(1+r^2)>0$ is in $\mathcal{B}$ with $\|f\|_{\infty}\le 1$.
% Proof:
% $\lim_{r\to0^+}f(r)=\frac{d}{dr}\log(1+r^2)|_{r=0}=0$ and $\lim_{r\to\infty}f(r)=0$, so $f\in\mathcal{B}$. $0<f(r)<1$ when $r>0$, because
% $g(r):=\log(1+r^2)-r$ has $g(0)=0$ and $g'(r)=\frac{2r}{1+r^2}-1=\frac{-(1-r)^2}{1+r^2}\le0$, so $g(r)<0$ on $r>0$.
It follows that the series $B(r):=\sum_{n=1}^{\infty}a_nf(a_nr)$ converges in $\mathcal{B}$ with $\|B\|_{\infty}\le a$ and hence $A(k)=|k|B(|k|)$ converges uniformly on compact subsets of $\R$ with $0\le A(k)\le a|k|$. All terms in the definition of $A(k)$ are even and strictly increasing on $k\ge0$, so the same is true for $A(k)$. The terms are non-negative and diverge as $k\to\infty$, so $\displaystyle{\lim_{k\to\infty}A(k)=\infty}$. Because $B\in\mathcal{B}$ we also have
$\displaystyle{\lim_{k\to\infty}\frac{A(k)}{|k|}=\lim_{k\to\infty}B(|k|)=0}$ and $\displaystyle{\lim_{k\to0}\frac{A(k)}{|k|}=\lim_{k\to0}B(|k|)=0}$.

For any $K>\kappa>0$ the series $\sum_{n=1}^{\infty}\frac{\log(1+a_n^2k^2)}{k^2}$ converges uniformly on $[\kappa,K]$ to $\frac{A(k)}{k^2}$, so we can integrate under the summation sign to obtain
\begin{align}
\int_{\kappa}^K\frac{A(k)}{k^2}\mathrm{d}k&=\sum_{n=1}^{\infty}\int_{\kappa}^K\frac{\log(1+a_n^2k^2)}{k^2}\mathrm{d}k\notag\\
&=\sum_{n=1}^{\infty}\left[\frac{-\log(1+a_n^2k^2)}{k}+2a_n\arctan(a_nk)\right]_{\kappa}^K\notag\\
&=\sum_{n=1}^{\infty}\frac{-\log(1+a_n^2K^2)}{K}+2a_n\arctan(a_nK)+\frac{\log(1+a_n^2\kappa^2)}{\kappa}-2a_n\arctan(a_n\kappa)\notag\\
&=-\frac{A(K)}{K}+\frac{A(\kappa)}{\kappa}+2\sum_{n=1}^{\infty}a_n(\arctan(a_nK)-\arctan(a_n\kappa))\,.\notag
\end{align}
In the limit $\kappa\to0$ and $K\to\infty$ the first two terms vanish. Because $\arctan$ is a montone function with $\arctan(0)=0$ and $\displaystyle{\lim_{r\to\infty}\arctan(r)=\frac{\pi}{2}}$ the series on the right-hand side converges to $a\frac{\pi}{2}$, so $\int_0^{\infty}\frac{A(k)}{k^2}\mathrm{d}k=\pi a$.
% Proof:
% Let $\epsilon>0$. Fix $N\in\mathbb{N}$ such that $0\le a-\sum_{n=1}^Na_n\le \epsilon$. Then
% \begin{align}
% 0&\le\sum_{n=N+1}^{\infty}a_n(\arctan(a_nK)-\arctan(a_n\kappa))\le \frac{\pi}{2}\epsilon\,.\notag
% \end{align}
% There is a $\delta>0$ such that for $\kappa\in(0,\delta)$ and $K\in(\delta^{-1},\infty)$ we have $\arctan(a_nK)\ge\frac{\pi}{2}-\frac{\epsilon}{N}$ and
% $\arctan(a_n\kappa)\le\frac{\epsilon}{N}$. Consequently,
% \begin{align}
% \left|\sum_{n=1}^Na_n(\arctan(a_nK)-\arctan(a_n\kappa))-a\frac{\pi}{2}\right|&\le
% \left|\sum_{n=1}^Na_n\left(\arctan(a_nK)-\frac{\pi}{2}-\arctan(a_n\kappa)\right)\right|+\epsilon\frac{\pi}{2}\notag\\
% &\le \sum_{n=1}^Na_n\frac{2\epsilon}{N}+\epsilon\frac{\pi}{2}\le \left(2a+\frac{\pi}{2}\right)\epsilon\,.\notag
% \end{align}
% The claimed limit follows.

Comparing the partial sums $A_N(k)=\sum_{n=1}^N\log(1+a_n^2k^2)$ and $\widehat{g_N}(k)$ (as defined in the proof of Proposition \ref{prop:defg}) we find from Lemma \ref{lem:etaprop} that for every $N\in\N$
\begin{align}
-A_N(\sqrt{\beta}k)&=\sum_{n=1}^N\log\left(\frac{1}{1+\beta a_n^2k^2}\right)\notag\\
&\le \sum_{n=1}^N\log(\hat{\eta}(a_nk))=\log(\widehat{g_N}(k))\notag\\
&\le \sum_{n=1}^N\log\left(\frac{1}{1+\alpha a_n^2k^2}\right)=-A_N(\sqrt{\alpha}k)\,.\notag
\end{align}
Exponentiating and taking the limit $N\to\infty$ proves the final estimate.
\end{proof*}

\subsection{Bounds for test functions of positive type in one dimension}\label{ssec:1dbounds}

It is known that the Fourier transform of a test-function $f\in C_0^{\infty}(\R)$ falls off faster than any power in $k$, because $(-\Delta)^nf\in C_0^{\infty}(\R)$ for all $n\in\N$ and hence $k^{2n}\hat{f}(k)$ is bounded on $\R$. However, if $f$ is non-zero, then $\hat{f}$ cannot fall off arbitrarily fast, because if
$|\hat{f}(k)|\le Ce^{-c|k|}$ for some $C,c>0$, then $f(x)$ extends to a holomorphic function $f(z)=\frac{1}{2\pi}\int e^{izk}\hat{f}(k)\mathrm{d}k$ on $\{z\in\C\mid |\mathrm{Im}(z)|< c\}$ and hence $f\equiv 0$. The following theorem characterises upper bounds for Fourier transforms of test-functions and is due to Ingham \cite{Ing1934} (slightly reformulated).
\begin{theorem}\label{thm:upper}
Let $l>0$ and $\epsilon:[0,\infty)\to(0,\infty)$ a monotonically decreasing measurable function. Then there is an $f\in C_0^{\infty}(\R)$ with support in $[-l,l]$ such that
\begin{align}
|\hat{f}(k)|&\le e^{-|k|\epsilon(|k|)}\label{eqn:upperbound}
\end{align}
for all $k\in\R$ if and only if
\begin{align}
\int_1^{\infty}\frac{\epsilon(k)}{k}\mathrm{d}k&<\infty\,.\label{eqn:intepsilon}
\end{align}
\end{theorem}

It is not hard to see that we can also require $\hat{f}\ge0$. Indeed, applying the theorem with $\frac12l$ and $\frac12\epsilon(k)$ instead of $l$ and $\epsilon(k)$ yields a function $g\in C_0^{\infty}(\R)$ with support in $\left[-\frac12l,\frac12l\right]$ such that $|\hat{g}(k)|\le e^{-\frac12|k|\epsilon(|k|)}$. We then choose $f=\tilde{g}*g\in C_0^{\infty}(\R)$ where $\tilde{g}(x)=\overline{g(-x)}$, so $f$ has support in $[-l,l]$ and $0\le\hat{f}(k)=|\hat{g}(k)|^2\le e^{-|k|\epsilon(|k|)}$ for all $k\in\R$. (See also Cor.~1.3.17 in \cite{Bjo1966}.)

\smallskip

Let us now consider lower bounds on $\hat{g}(k)$. When $g$ is constructed as in Proposition \ref{prop:defg}, we see from Lemma \ref{lem:defA} that $\hat{g}(k)\ge e^{-A(\sqrt{\beta}k)}\ge e^{-\sqrt{\beta}a|k|}$. The following result shows that we can often get a better lower bound, which is of the same form as the upper bound in Theorem \ref{thm:upper} when $\epsilon(k)$ satisfies an additional assumption.

\begin{theorem}\label{thm:lower}
Let $l>0$, $\gamma\in(0,1)$ and $\epsilon:[0,\infty)\to(0,\infty)$ a decreasing measurable function satisfying (\ref{eqn:intepsilon}) and such that
$\displaystyle{\lim\inf_{k\to\infty}k^{\gamma}\epsilon(k)>0}$. Then there is a non-negative even $g\in C_0^{\infty}(\R,\R)$ with support in $[-l,l]$ such that
$\int g=1$ and
\begin{align}
\hat{g}(k)&\ge e^{-|k|\epsilon(|k|)}\label{eqn:lowerbound}
\end{align}
for all $k\in\R$.
\end{theorem}
% Note that for this lower bound it actually suffices to replace the assumed estimate (\ref{eqn:intepsilon}) by the weaker estimate
% \begin{align}
% \int_1^{\infty}\frac{\epsilon(k)^{\frac{1}{1-\gamma}}}{k}\mathrm{d}k&<\infty\,.\label{eqn:intepsilon}
% \end{align}

\begin{proof*}
We will find a sequence $\{a_n\}_{n\in\N}$ in $(0,\infty)$ such that $a:=\sum_{n=1}^{\infty}a_n\le l$ and such that the function $A(k)$ of Lemma \ref{lem:defA} satisfies $A(\sqrt{\beta}k)\le k\epsilon(k)$ for all $k\ge0$. The result then follows from Proposition \ref{prop:defg} and Lemma \ref{lem:defA}.

We first define the function $\epsilon':(0,\infty)\to[0,\infty)$ by
\begin{align}
\epsilon'(k)&:=\frac{1}{k^{\gamma}}\inf_{l\ge k}l^{\gamma}\epsilon(l)\,.\notag
\end{align}
Note that $0\le \epsilon'(k)\le\epsilon(k)$ and that the inequality (\ref{eqn:intepsilon}) also holds for $\epsilon'$ instead of $\epsilon$. If $k'\ge k>0$ we have
$\displaystyle{\epsilon'(k)=\frac{1}{k^{\gamma}}\min\left\{\inf_{l\in [k,k']}l^{\gamma}\epsilon(l),\inf_{l\ge k'}l^{\gamma}\epsilon(l)\right\}}$ and
\begin{align}
\frac{1}{k^{\gamma}}\inf_{l\ge k'}l^{\gamma}\epsilon(l)&\ge \frac{1}{(k')^{\gamma}}\inf_{l\ge k'}l^{\gamma}\epsilon(l)
=\epsilon'(k')\,,\notag\\
\frac{1}{k^{\gamma}}\inf_{l\in [k,k']}l^{\gamma}\epsilon(l)&\ge \frac{1}{k^{\gamma}}\left(k^{\gamma}\epsilon(k')\right)
=\epsilon(k')\ge \epsilon'(k')\,,\notag
\end{align}
so $\epsilon'(k)\ge \epsilon'(k')$, i.e.~$\epsilon'$ is also monotonically decreasing. Because $\displaystyle{\lim\inf_{k\to\infty}k^{\gamma}\epsilon(k)>0}$ by assumption, this also entails that $\epsilon'(k)>0$ for all $k>0$.
% Otherwise we would have $\epsilon'(k_0)=0$ for some $k_0>0$ and hence $\epsilon'(k)=0$ for all $k\ge k_0$, which would yield lead to the contradictory
% statement $\displaystyle{\lim\inf_{k\to\infty}k^{\gamma}\epsilon(k)=0}$.
We can extend $\epsilon'(k)$ to $k=0$ by setting $\displaystyle{\epsilon'(0):=\lim_{k\to0^+}\epsilon'(k)}$, where the limit exists and is at most $\epsilon(0)$, because $\epsilon'$ is decreasing and $\epsilon'(k)\le\epsilon(k)\le\epsilon(0)$. Finally, $k\mapsto k^{\gamma}\epsilon'(k)$ is monotonically increasing (a property which $\epsilon$ need not have).

It now suffices to prove that $A(\sqrt{\beta}k)\le k\epsilon'(k)$ for all $k>0$, because $\epsilon'(k)\le \epsilon(k)$ and at $k=0$ we have $A(0)=0$. Let $\delta=\frac{1}{1-\gamma}$. We see from (\ref{eqn:intepsilon}) that
\begin{align}
a'&:=\sum_{n=1}^{\infty}\frac{\epsilon'(n)^{\delta}}{n}\le \epsilon'(1)^{\delta}+\int_1^{\infty}\frac{\epsilon'(k)^{\delta}}{k}\mathrm{d}k
\le \epsilon'(1)^{\delta}+\epsilon'(1)^{\delta-1}\int_1^{\infty}\frac{\epsilon'(k)}{k}\mathrm{d}k<\infty\,,\notag
\end{align}
because $\epsilon'$ is decreasing. We now set $a_n:=\lambda\frac{\epsilon'(n)^{\delta}}{n}$ for $n\ge1$, where
\begin{align}
\lambda&=\min\left\{\frac{l}{a'},
\left( \pi\sqrt{\beta}\epsilon(0)^{\delta-1} + \frac{4}{(1-\gamma^2)}\beta^{\frac{1}{2\delta}} \right)^{-\delta}\right\}
\label{eqn:deflambda}
\end{align}
with $\beta=\frac{7}{40}$. Because $\sum_{n=1}^{\infty}a_n\le l$, the function $g$ of Proposition \ref{prop:defg} has the desired support property. Also note that $\lambda\le 1$, because $\frac{4}{1-\gamma^2}\beta^{\frac{1}{2\delta}}\ge 1$ for all $\gamma\in(0,1)$. We now use the fact that $n\mapsto\frac{\epsilon'(n)^{\delta}}{n}$ is decreasing to estimate for all $k\ge0$
\begin{align}
A(\sqrt{\beta}k)&=\sum_{n=1}^{\infty}\log\left(1+\beta\lambda^2k^2\frac{\epsilon'(n)^{2\delta}}{n^2}\right)
\le\int_0^{\infty}\log\left(1+\beta\lambda^2k^2\frac{\epsilon'(r)^{2\delta}}{r^2}\right)\mathrm{d}r\,.\label{eqn:estA1}
\end{align}
We split the integral and estimate
\begin{align}
\int_k^{\infty}\log\left(1+\beta\lambda^2k^2\frac{\epsilon'(r)^{2\delta}}{r^2}\right)\mathrm{d}r&\le
\int_k^{\infty}\log\left(1+\beta\lambda^2k^2\frac{\epsilon'(k)^{2\delta}}{r^2}\right)\mathrm{d}r\notag\\
&\le\sqrt{\beta}\lambda k\epsilon'(k)^{\delta}\int_0^{\infty}\log\left(1+\frac{1}{r^2}\right)\mathrm{d}r\notag\\
&\le\lambda^{\frac{1}{\delta}}\pi\sqrt{\beta}\epsilon(0)^{\delta-1}k\epsilon'(k)\,,\label{eqn:estA2}
\end{align}
where we used $\epsilon'(k)^{\delta}=\epsilon'(k)^{\delta-1}\epsilon'(k)\le \epsilon'(0)^{\delta-1}\epsilon'(k)\le \epsilon(0)^{\delta-1}\epsilon'(k)$,
$\lambda\le\lambda^{\frac{1}{\delta}}$ and the integral 
\begin{align}
\int_0^{\infty}\log\left(1+\frac{1}{r^2}\right)\mathrm{d}r&=\left[r\log\left(1+\frac{1}{r^2}\right)+2\arctan(r)\right]_0^{\infty}=\pi\,.\notag
\end{align}
Using the fact that $k^{\gamma}\epsilon'(k)$ is increasing and $2(\delta-1)=2\gamma\delta$ we also have
\begin{align}
\int_0^k\log\left(1+\beta\lambda^2k^2\frac{\epsilon'(r)^{2\delta}}{r^2}\right)\mathrm{d}r&
=\int_0^k\log\left(1+\beta\lambda^2k^2\frac{r^{2\gamma\delta}\epsilon'(r)^{2\delta}}{r^{2\delta}}\right)\mathrm{d}r\notag\\
&\le\int_0^k\log\left(1+\beta\lambda^2k^2\frac{k^{2\gamma\delta}\epsilon'(k)^{2\delta}}{r^{2\delta}}\right)\mathrm{d}r\notag\\
&=\int_0^k\log\left(1+\beta\lambda^2k^{2\delta}\frac{\epsilon'(k)^{2\delta}}{r^{2\delta}}\right)\mathrm{d}r\notag\\
&\le\beta^{\frac{1}{2\delta}}\lambda^{\frac{1}{\delta}}k\epsilon'(k)\int_0^{\infty}\log\left(1+\frac{1}{r^{2\delta}}\right)\mathrm{d}r\notag\\
&\le\lambda^{\frac{1}{\delta}}\frac{4}{1-\gamma^2}\beta^{\frac{1}{2\delta}}k\epsilon'(k)\,,\label{eqn:estA3}
\end{align}
where we estimated the integral
\begin{align}
\int_0^{\infty}\log(1+r^{-2\delta})\mathrm{d}r&=\left[r\log(1+r^{-2\delta})\right]_0^{\infty}
-\int_0^{\infty}r\frac{-2\delta r^{-2\delta-1}}{1+r^{-2\delta}}\mathrm{d}r\notag\\
&=0+2\delta\int_0^{\infty}\frac{1}{1+r^{2\delta}}\mathrm{d}r\notag\\
&\le 2\delta\int_0^11\mathrm{d}r
+2\delta\int_1^{\infty}\frac{1}{r^{2\delta}}\mathrm{d}r\notag\\
&=2\delta-2\delta\frac{1}{1-2\delta}=\frac{2}{1-\gamma}+\frac{2}{1+\gamma}=\frac{4}{1-\gamma^2}\,.\notag
\end{align}
% To see why the boundary terms vanish we use the substitution $r=\frac{1}{s}$ to note that
% \begin{align}
% \lim_{r\to\infty}r\log\left(1+r^{-2\delta}\right)&=\lim_{s\to 0^+}\frac{\log\left(1+s^{2\delta}\right)}{s}\notag\\
% &=\partial_s\log\left(1+s^{2\delta}|_{s=0}=\frac{2\delta s^{2\delta-1}}{1+s^{2\delta}}|_{s=0}=0\,,\notag
% \lim_{r\to0^+}r\log\left(1+r^{-2\delta}\right)&=\lim_{s\to\infty}\frac{\log\left(1+s^{2\delta}\right)}{s}=0\,.\notag
% \end{align}
Combining (\ref{eqn:estA1}), (\ref{eqn:estA2}), (\ref{eqn:estA3}) and (\ref{eqn:deflambda}) we find that $A(\sqrt{\beta}k)\le |k|\epsilon(|k|)$ for all $k\in\R$.
\end{proof*}

To understand better what kind of fall-off properties can be achieved by the construction of Proposition \ref{prop:defg}, it would be desirable to get a better understanding of the class of functions $A(r)$ appearing in Lemma \ref{lem:defA}. Instead of pursuing this question, we will consider an example.

\begin{example}\label{ex:Gevrey1}
For a lower bound of Gevrey type it is possible to give a more explicit construction. Let $\gamma\in(0,1)$ and $c,l>0$ be given and let $\beta=\frac{7}{40}$ as in Lemma \ref{lem:defA}. Furthermore, let $\{a_n\}_{n\in\N}$ be any positive sequence such that $\sum_{n=1}^{\infty}a_n^{\gamma}$ converges, 
$a=\sum_{n=1}^{\infty}a_n<l$ and
\begin{align}
\sum_{n=1}^{\infty}a_n^{\gamma}&\le c\left(\frac{2-\gamma}{\gamma}\right)^{\frac{\gamma-2}{2}}\beta^{-\frac{\gamma}{2}}\,.\notag
\end{align}
Then the test-function $g$ constructed in Proposition \ref{prop:defg} is positive, even, of positive type and supported in $[-l,l]$. In addition we will now establish the lower bound
\begin{align}
\hat{g}(k)&\ge e^{-c|k|^{\gamma}}\notag
\end{align}
for all $k\in\R$. For this purpose we first note that the function $h(k)=\frac{k^{2-\gamma}}{1+k^2}$ on $k\ge0$ is non-negative with
$h'(k)=\frac{k^{1-\gamma}}{(1+k^2)^2}(2-\gamma-\gamma k^2)$. It follows that $h$ has a global maximum at the unique critical point $k_1=\sqrt{\frac{2-\gamma}{\gamma}}$, where $h(k_1)=\frac{\gamma}{2}\left(\frac{2-\gamma}{\gamma}\right)^{\frac{2-\gamma}{2}}$. Furthermore, the function $f(k)=k^{-\gamma}\log(1+k^2)$ on $k>0$ is non-negative and has $f'(k)=k^{-\gamma-1}\left(\frac{2k^2}{1+k^2}-\gamma\log(1+k^2)\right)$. It follows that $f'(0)=0$ and one can show that $f'$ has one further root $k_*>0$, which may be expressed in terms of the Lambert $W$ function, but we refrain from doing so here.
% The roots for $k\ge0$ are
% \begin{align}
% k&=\sqrt{x\frac{1}{W_j\left(xe^x\right)}-1}\notag
% \end{align}
% where $x=\frac{-2}{\gamma}$ and $j\in\{0,-1\}$.
Instead we note that at $k_*$ the function $f$ has a global maximum and we have $\log(1+k_*^2)=\frac{2k_*^2}{\gamma(1+k_*^2)}$. It follows that for all $k\ge0$
\begin{align}
f(k)&\le f(k_*)=k_*^{-\gamma}\frac{2}{\gamma}\cdot\frac{k_*^2}{1+k_*^2}=\frac{2}{\gamma}h(k_*)\le \frac{2}{\gamma}h(k_1)
=\left(\frac{2-\gamma}{\gamma}\right)^{\frac{2-\gamma}{2}}\,.\notag
\end{align}
We rewrite this inequality in the form $\log(1+k^2)\le\left(\frac{2-\gamma}{\gamma}\right)^{\frac{2-\gamma}{2}}|k|^{\gamma}$ for $k\in\R$ to estimate the function $A$ of Lemma \ref{lem:defA},
\begin{align}
A(\sqrt{\beta}k)&=\sum_{n=1}^{\infty}\log(1+\beta a_n^2k^2)\le 
\left(\frac{2-\gamma}{\gamma}\right)^{\frac{2-\gamma}{2}}\beta^{\frac{\gamma}{2}}\left(\sum_{n=1}^{\infty}a_n^{\gamma}\right)|k|^{\gamma}
\le c|k|^{\gamma}\notag
\end{align}
by our assumptions on the sequence $\{a_n\}_{n\in\N}$. It now follows immediately from Lemma \ref{lem:defA} that
$\hat{g}(k)\ge e^{-A(\sqrt{\beta}k)}\ge e^{-c|k|^{\gamma}}$, as desired.
\end{example}

\subsection{Test functions of positive type in higher dimensions}\label{ssec:higherdbounds}

We now consider $\R^d$ for $d\in\N$, using the Euclidean norm $|x|$.

First we generalise the construction of Proposition \ref{prop:defg} to $\R^d$. Let $\{a_n\}_{n\in\N}$ be a sequence in $(0,\infty)$ such that the series $a=\sum_{n=1}^{\infty}a_n$ converges and $g\in C_0^{\infty}(\R)$ the corresponding function of Proposition \ref{prop:defg}. For $d\in\N$ we define the functions $f_0,f\in C_0^{\infty}(\R^d)$ by $f_0(x):=g(x^1)\cdots g(x^d)$ and
\begin{align}
f(x)&:=\int_{O(d)} f_0(Rx)\,\mathrm{d}\mu(R)\,,\label{eqn:deff}
\end{align}
where $R\in O(d)$ is a rotation matrix and $\mathrm{d}\mu(R)$ the Haar measure on the Lie group $O(d)$ of real orthogonal $d\times d$-matrices with total volume $1$.
\begin{proposition}\label{prop:deff}
Let $\{a_n\}_{n\in\N}$ be a decreasing sequence in $(0,\infty)$ such that the series $a=\sum_{n=1}^{\infty}a_n$ converges and $f$ the function defined in 
(\ref{eqn:deff}) for some $d\in\N$. Then $f\in C_0^{\infty}(\R^d)$ is smooth, non-negative, rotation invariant, supported in the closed ball $B_{\sqrt{d}a}=\{x\in\R^d|\ |x|\le\sqrt{d}a\}$ and of positive type with $\int f=1$. Furthermore, for all $l\in\N$ we have the following estimates on $L^1$ and suppremum norms:
\begin{align}
\|\partial_1^lf\|_1&\le\frac{(3\sqrt{d})^l}{a_1\cdots a_l}\notag\\
\|\partial_1^lf\|_{\infty}&\le \left(\frac{3}{2a_1}\right)^d\frac{(3\sqrt{d})^l}{a_1\cdots a_l}\,.\notag
\end{align}
\end{proposition}
\begin{proof*}
Using the properties of the function $g$ defined in Proposition \ref{prop:defg} it follows that $f_0=g^{\otimes d}$ is in $C_0^{\infty}(\R^d)$ with support in $B_{\sqrt{d}a}$. Moreover, $f_0$ is even with $f_0\ge0$ and $\int f_0=1$. As $\widehat{f_0}=\hat{g}^{\otimes d}\ge0$, $f_0$ is of positive type. Furthermore, for any multiindex $\alpha\in\N_0^d$ we have
\begin{align}
\|\partial^{\alpha}f_0\|_1&=\prod_{j=1}^d\|\partial_j^{\alpha_j}g\|_1
\le\prod_{j=1}^d\frac{3^{\alpha_j}}{a_1\cdots a_{\alpha_j}}
\le \frac{3^{|\alpha|}}{a_1\cdots a_{|\alpha|}}\,,\notag
\end{align}
where we used the fact that the sequence $\{a_n\}_{n\in\N}$ is decreasing. Similarly,
\begin{align}
\|\partial^{\alpha}f_0\|_{\infty}&\le\prod_{j=1}^d\|\partial_j^{\alpha_j}g\|_{\infty}
\le\prod_{j=1}^d\frac{3}{2a_1}\frac{3^{\alpha_j}}{a_1\cdots a_{\alpha_j}}
\le \left(\frac{3}{2a_1}\right)^d\frac{3^{|\alpha|}}{a_1\cdots a_{|\alpha|}}\,.\notag
\end{align}
For any unit vector $v\in\R^d$ with components $v^i$ we have $\sum_{j=1}^d|v^i|\le \sqrt{d}|v|=\sqrt{d}$ by the Cauchy-Schwarz Inequality and hence
\begin{align}
\|(v^i\partial_i)^lf_0\|_1&\le\sum_{|\alpha|=l}\frac{l!}{\alpha!}|v^{\alpha}|\cdot\|\partial^{\alpha}f_0\|_1
\le \frac{3^l}{a_1\cdots a_l}\sum_{|\alpha|=l}\frac{l!}{\alpha!}|v^{\alpha}|
=\frac{3^l}{a_1\cdots a_l}\left(\sum_{j=1}^d|v^i|\right)^l\le \frac{(3\sqrt{d})^l}{a_1\cdots a_l}\notag
\end{align}
and similarly
\begin{align}
\|(v^i\partial_i)^lf_0\|_{\infty}&\le\sum_{|\alpha|=l}\frac{l!}{\alpha!}|v^{\alpha}|\cdot \|\partial^{\alpha}f_0\|_{\infty}
\le \left(\frac{3}{2a_1}\right)^d\frac{3^l}{a_1\cdots a_l}\sum_{|\alpha|=l}\frac{l!}{\alpha!}|v^{\alpha}|\le \left(\frac{3}{2a_1}\right)^d\frac{(3\sqrt{d})^l}{a_1\cdots a_l}\,.\notag
\end{align}
It follows that $f_0$ satisfies all desired properties, except for the rotation invariance.

For any rotation matrix $R$ we note that the function $f_0(Rx)$ is still non-negative with integral $1$ and support in the ball $B_{\sqrt{d}a}$ and that its Fourier transform is $\widehat{f_0}((R^{-1})^Tk)\ge0$. It follows that $f(x)$ as defined in (\ref{eqn:deff}) is rotationally invariant, non-negative, smooth (because we integrate over a compact group) with $\int f=1$ and support in the ball $B_{\sqrt{d}a}$ and of positive type. Writing $e_1$ for the unit vector in the $x^1$ direction we have
\begin{align}
\partial_1^lf(x)&=\int_{O(d)} ((Re_1)^i\partial_i^lf_0)(Rx)\,\mathrm{d}\mu(R)\notag
\end{align}
and since $Re_1\in R^d$ is a unit vector the estimates for $\partial_1^lf$ follow from those for $f_0$.
\end{proof*}

\begin{example}\label{ex:Gevrey2}
Using the sequence $a_n=R\frac{\rho-1}{\rho}n^{-\rho}$ of Example \ref{ex:Gevrey0} in Proposition \ref{prop:deff} yields a non-negative rotation invariant test-function $f\in C_0^{\infty}(\R^d)$ of positive type with support in the ball of radius $\sqrt{d}R$ and with 
\begin{align}
\|\partial_1^lf\|_1&\le\left(\frac{3\sqrt{d}\rho}{R(\rho-1)}\right)^ll!^{\rho}\notag
\end{align}
for any $l\in\N$. A similar estimate holds for partial derivatives in all directions, so it follows that $f$ is in the Gevrey class of order $\rho>1$ and therefore
\begin{align}
|\hat{f}(k)|&\le c^{-1}e^{-c|k|^{\frac{1}{\rho}}}\notag
\end{align}
for some $c>0$ and all $k\in\R^d$.
\end{example}

\section{Separable states in relativistic QFT}\label{sec:QFT}

We now turn to the task of constructing separable states in QFT. We will consider a free, real scalar field of mass $m>0$ and for simplicity we will first focus on Minkowski spacetime $M$ of dimension $1+d\ge2$ with inertial coordinates $\mathbf{x}=(x^0,x)$ and spacetime signature $(+-\ldots-)$. We will use Planck units ($c=\hbar=G=k_B=1$).

A classical scalar field is a real-valued solution $\varphi\in C^{\infty}(M)$ satisfying the Klein-Gordon equation $(\Box+m^2)\varphi=0$, where $\Box=\partial_0^2-\Delta=\partial_0^2-\sum_{j=1}^d\partial_j^2$ is the wave operator. There are unique advanced ($-$) and retarded ($+$) fundamental solutions $E^{\pm}:C_0^{\infty}(M)\to C^{\infty}(M)$ such that $(\Box+m^2)E^{\pm}f=f$ and $\mathrm{supp}(E^{\pm}f)\subseteq J^{\pm}(\mathrm{supp}f)$ for all $f\in C_0^{\infty}(M)$, where $J^{\pm}$ denotes the causal future ($+$) or past ($-$). With these fundamental solutions one can easily construct solutions to the Klein-Gordon equation of the form $\varphi=Ef$ where $E:=E^--E^+$ and $f\in C_0^{\infty}(M)$.

The corresponding quantum theory is described by a $^{\ast}$-algebra $\mathcal{A}$ which is generated by a unit $I$ and elements $\phi(f)$ with $f\in C_0^{\infty}(M)$, satisfying the relations
\begin{enumerate}[(i)]
\item $f\mapsto \phi(f)$ is complex linear,
\item $\phi(f)^{\ast}=\phi(\bar{f})$,
\item $\phi((\Box+m^2)f)=0$,
\item $[\phi(f),\phi(g)]=iE(f,g)I$
\end{enumerate}
for all $f,g\in C_0^{\infty}(M)$, where $E(f,g):=\int_M fEg$ encodes the canonical commutation relation of the theory. The property (iii) implements the equation of motion $(\Box+m^2)\phi=0$ in a distributional sense.

A quantum state is a linear functional $\omega:\A\to\C$ which is positive, $\omega(A^{\ast}A)\ge0$ for all $A\in\A$, and normalised, $\omega(I)=1$. We will focus on quasi-free states, which are determined completely by a two-point distribution $\omega_2$ on $M^{\times 2}$ such that
\begin{align}
\omega_2(f,g)&=\omega(\phi(f)\phi(g))\label{def:2pt}
\end{align}
for all $f,g\in C_0^{\infty}(M)$. For such states, expectation values of other operators can be determined from the power series
\begin{align}\label{eqn:qfstate}
\sum_{n=0}^{\infty}\frac{i^n}{n!}\omega(\phi(f)^n)&=e^{-\frac12\omega_2(f,f)}
\end{align}
for all real-valued $f\in C_0^{\infty}(M)$ and from the algebraic relations of $\A$. 

In order to be the two-point distribution of a quasi-free state, a distribution $\omega_2$ on $M^{\times 2}$ must satisfy the following three conditions:
\begin{enumerate}
\item[(C1)] $\omega_2(\mathbf{x},\mathbf{x}')$ is a (distributional) solution to the Klein-Gordon equation in each variable,
\item[(C2)] the anti-symmetric part of $\omega_2$ is given by the canonical commutation relation,
\begin{align}
\omega_{2-}(f,g)&:=\frac12\omega_2(f,g)-\frac12\omega_2(g,f)=\frac12iE(f,g)\,.\label{eqn:commutator2pt}
\end{align}
\item[(C3)] $\omega_2$ must satisfy
\begin{align}
\omega_2(\bar{f},f)&=\omega(\phi(f)^*\phi(f))\ge0\label{eqn:positive2pt}
\end{align}
for all $f\in C_0^{\infty}(M)$, or, equivalently (by the Cauchy-Schwarz Inequality), the symmetric part $\omega_{2+}(f,g):=\frac12\omega_2(f,g)+\frac12\omega_2(g,f)$ of $\omega_2$ must satisfy $\omega_{2+}(f,f)\omega_{2+}(g,g)\ge |\omega_{2-}(f,g)|^2$ for all real-valued $f,g\in C_0^{\infty}(M)$.
\end{enumerate}

$\omega_{2+}$ is not fixed in advance by the theory and different choices of $\omega_{2+}$ correspond to different quasi-free states, of which there are many, including the Minkowski vacuum state $\omega^{(0)}$, KMS states that describe a thermal equilibrium and many others. A quasi-free state $\omega$ is called Hadamard iff $\omega_2-\omega_2^{(0)}$ is smooth. (Note that this definition depends on the choice of the mass parameter.) Hadamard states are of crucial importance in the perturbative treatment of self-interactions, cf.~\cite{HW2015}, and we note that all KMS states are Hadamard \cite{SV2000}.

A key property of Hadamard states is that the expected energy density of the quantum field (after appropriate renormalisation) is a smoothly varying function on spacetime. The energy density of a classical solution $\varphi$ to the Klein-Gordon equation of mass $m$ (for the given choice of intertial time)
is\footnote{This expression is the $00$-component of the stress-energy-momentum tensor, which can be derived by varying the Lagrangian density $\frac12(-|\nabla\varphi|^2+m^2\varphi^2)\mathrm{d}vol_g$ w.r.t.~the metric field $g^{\mu\nu}$. Here $\mathrm{d}vol_g$ is the metric volume form and we assume the field has minimal coupling to the metric, i.e.~the Lagrangian density has no term $\frac12\xi R$ where $R$ is the scalar curvature.}
\begin{align}
E[\varphi](\mathbf{x})&:=\frac12\left(\sum_{\mu=0}^d(\partial_{\mu}\varphi)^2(\mathbf{x})+m^2\varphi^2(\mathbf{x})\right)\,.\notag
\end{align}
For the quantum field the expected normal ordered energy density in any Hadamard state $\omega$ is given by\footnote{Because we will only consider the energy density in Minkowski spacetime, we refrain from applying the generally covariant Hadamard renormalisation scheme \cite{HW2015}. The latter might yield additional terms in the energy density when $m>0$, which correspond to the expected energy density of the Minkowski vacuum.}
\begin{align}
\omega(E^{\mathrm{ren}}(\mathbf{x}))&:=\frac12\left(\sum_{\mu=0}^d\partial_{\mu}\partial_{\mu}'+m^2\right)(\omega_2-\omega^{(0)}_2)
(\mathbf{x},\mathbf{x}')|_{\mathbf{x}'=\mathbf{x}}\,.\label{eqn:Eren}
\end{align}

Starting with Minkowski spacetime we will now review the relation between quasi-free states and the positive type of distributions. We will then consider the prevalence of entanglement in quantum states and apply the results from Section \ref{sec:positivity} to construct separable states.

\subsection{Distributions of positive type and homogeneous quantum states}\label{ssec:positivitystates}

A quasi-free state is homogeneous, i.e.~translation invariant in spacetime, iff its two-point distribution is of the form
\begin{align}
\omega_2(\mathbf{x},\mathbf{x}')&=\lambda(\mathbf{x}-\mathbf{x}')\notag
\end{align}
for some distribution $\lambda$ on $M$. The positivity of the state $\omega$, in the form of condition (C3), entails that
\begin{align}
\lambda(\bar{f}*\tilde{f})&\ge0\,,\notag
\end{align}
where we recall that $\tilde{f}(\mathbf{x})=f(-\mathbf{x})$ and $*$ is the convolution. I.e., $\lambda$ is a distribution of positive type and in particular it is tempered with $\widehat{\lambda}\ge0$ (cf.~\cite{RS1980} Thm.~IX.10). $\omega_{2\pm}$ correspond to the even and odd parts of $\lambda$, which we will also denote by $\lambda_{\pm}$.
% $\omega_{2\pm}(\mathbf{x},\mathbf{y})=\frac12\lambda(\mathbf{x}-\mathbf{y})\pm\frac12\lambda(\mathbf{y}-\mathbf{x})$.

The prime example of a homogeneous state is the Minkowski vacuum state $\omega^{(0)}$, which is a quasi-free state with
\begin{align}\label{eqn:vacuum}
\omega^{(0)}_2(\mathbf{x},\mathbf{x}')&=(2\pi)^{-d}\int e^{-ik_0((x')^0-x^0)+ik\cdot(x'-x)}\theta(k_0)\delta(k_0^2-|k|^2-m^2)\mathrm{d}k_0\mathrm{d}^dk\\
&=(2\pi)^{-d}\int e^{-i\nu(k)((x')^0-x^0)+ik\cdot(x'-x)}\frac{1}{2\nu(k)}\mathrm{d}^dk\,,\notag
\end{align}
where $\theta$ is the Heaviside step function and $\nu(k):=\sqrt{|k|^2+m^2}$. Here we recognise that the Fourier transform is a positive measure $\widehat{\lambda^{(0)}}(\mathbf{k})=\theta(k_0)\delta(k_0^2-|k|^2-m^2)$. We can find the canonical commutator function from
\begin{align}
\frac{i}{2}E(\mathbf{x},\mathbf{x}')&=\omega^{(0)}_{2-}(\mathbf{x},\mathbf{x}')=
\frac12(2\pi)^{-d}\int e^{-ik_0((x')^0-x^0)+ik\cdot(x'-x)}\mathrm{sign}(k_0)\delta(k_0^2-|k|^2-m^2)\mathrm{d}k_0\mathrm{d}^dk\,,\notag
\end{align}
which consists of a positive measure supported on the positive mass shell $k_0=\nu(k)$ and a negative measure supported on the negative mass shell $k_0=-\nu(k)$.

A general homogeneous quasi-free state is determined by a distribution $\lambda$ such that $\widehat{\lambda}\ge0$ is supported on the mass shells $k_0^2=\nu(k)^2$ and which has odd part $\lambda_-=\lambda^{(0)}_-$ (see conditions (C1-3) above). In particular, the even part $\lambda_+$ needs to compensate for the negative measure of $\lambda_-$ on the negative mass shell. It follows that $\widehat{\lambda}\ge\widehat{\lambda^{(0)}}$, or, equivalently, $\lambda-\lambda^{(0)}$ is of positive type. If the state is also required to be Hadamard, then $l:=\lambda-\lambda^{(0)}$ must be a smooth function of positive type. This is known to imply that $\widehat{l}$ is a positive, finite measure by Bochner's theorem (cf.~\cite{RS1980} Thm.~IX.9).

It is often convenient to express two-point distributions in terms of their initial data at $x^0=0$. We will briefly review this for general two-point distributions before we restrict attention to homogeneous states.

We let $\Sigma=\{(0,x)\}$ be the $x^0=0$ hyperplane. For a general two-point distribution $\omega_2$ on $M\times M$ we define the (distributional) initial data on $\Sigma\times\Sigma$ by
\begin{align}\label{eqn:data}
\omega_{2,00}&:=\omega_2|_{\Sigma\times\Sigma}\,,\nonumber\\
\omega_{2,10}&:=(\partial_0\otimes 1)\omega_2|_{\Sigma\times\Sigma}\,,\nonumber\\
\omega_{2,01}&:=(1\otimes \partial_0')\omega_2|_{\Sigma\times\Sigma}\,,\nonumber\\
\omega_{2,11}&:=(\partial_0\otimes\partial_0')\omega_2|_{\Sigma\times\Sigma}\,,
\end{align}
where we note that these restrictions are defined by a microlocal argument, because $\omega_2$ is a bi-solution to the Klein-Gordon equation, so its wave front set only contains light-like covectors. We use a similar notation for the initial data of $\omega_{2\pm}$. By the canonical commutation relations we have $\omega_{2-}=\frac{i}{2}E$ and hence
\begin{align}\label{eqn:Edata}
\omega_{2-,00}(x,x')&=\omega_{2-,11}(x,x')=0\,,\nonumber\\
\omega_{2-,10}(x,x')&=-\omega_{2-,01}(x,x')=\frac{i}{2}\delta(x'-x)\,.
\end{align}
Any distributionial initial data $\omega_{2,ij}$ define a bi-solution to the Klein-Gordon equation, so condition (C1) is automatically satisfied. To guarantee condition (C2), $\omega_{2,00}$ and $\omega_{2,11}$ must be symmetric in their arguments and $\omega_{2,10}$ and $\omega_{2,01}$ must differ from $\pm i\frac12\delta$ by a symmetric contribution. The positivity condition (C3) requires that the matrix
$\begin{pmatrix} \omega_{2,00}& \omega_{2,01}\\   \omega_{2,10}& \omega_{2,11}\end{pmatrix}$,
viewed as an integral kernel acting on $C_0^{\infty}(\Sigma,\C^{\oplus 2})$, is non-negative.

These expressions simplify in the case of a homogeneous state. Indeed, $l=\lambda-\lambda^{(0)}$ is a real-valued solution to the Klein-Gordon equation, which is determined by only two pieces of initial data, 
\begin{align}\label{eqn:dataTinv}
l_0&:=l|_{\Sigma}\,,\nonumber\\
l_1&:=\partial_0l|_{\Sigma}\,.
\end{align}
The canonical commutation relations imply that $l_0$ is a real and even distribution and $l_1$ is real and odd. The corresponding two-point distribution has initial data
\begin{align}
\omega_{2,00}(x,x')&=\omega^{(0)}_{2,00}(x,x')+l_0(x-x')\,,\nonumber\\
\omega_{2,10}(x,x')&=\frac{i}{2}\delta(x-x')+l_1(x-x')\,,\nonumber\\
\omega_{2,01}(x,x')&=-\frac{i}{2}\delta(x-x')-l_1(x-x')\,,\nonumber\\
\omega_{2,11}(x,x')&=\omega^{(0)}_{2,11}(x,x')+(-\Delta+m^2)l_0(x-x')\,.\notag
\end{align}
Because $l$ is of positive type, it has a Fourier transform and so do its initial data. We have
$\widehat{l}(\mathbf{k})=\pi(\delta(k_0-\nu(k))+\delta(k_0+\nu(k)))\left(\widehat{l_0}(k)+\frac{\widehat{l_1}(k)}{ik_0}\right)$ where we recall that $\nu(k)=\sqrt{|k|^2+m^2}$. From the positive type of $l$ we find that
\begin{align}
\left|\widehat{l_1}(k)\right|&\le \nu(k)\widehat{l_0}(k)\,.\notag
\end{align}
To conclude this secion we note that the initial values of the Minkowski vacuum state are given by
\begin{align}
\omega_{2,00}^0(x,x')&=\frac{1}{2(2\pi)^d}\int e^{i(x-x')\cdot k}\frac{1}{\nu(k)}\mathrm{d}^dk
=c_d|x-x'|^{\frac{1-d}{2}}K_{\frac{d-1}{2}}(m|x-x'|)\,,\label{eqn:vacuumvalues}
\end{align}
when $x\not=x'$, where $c_d>0$ is a constant and $K_{\nu}$ is a modified Bessel function.
% Proof:
% For $\hat{w}(k)=\frac12(m^2+|k|^2)^{-\frac12}$ we have
% \begin{align}
% -k\cdot\nabla_k\hat{w}(k)&=-\frac12m^2(m^2+|k|^2)^{-\frac32}+\hat{w}(k)\notag\\
% (-k\cdot\nabla_k)^2\hat{w}(k)&=\frac32m^4(m^2+|k|^2)^{-\frac52}-2m^2(m^2+|k|^2)^{-\frac32}+\hat{w}(k)\notag\\
% \Delta_p\hat{w}(k)&=-\frac12\nabla_k\cdot k(m^2+|k|^2)^{-\frac32}\notag\\
% &=\frac32|k|^2(m^2+|k|^2)^{-\frac52}-\frac12d(m^2+|k|^2)^{-\frac32}\notag\\
% &=-\frac32m^2(m^2+|k|^2)^{-\frac52}-\frac12(d-3)(m^2+|k|^2)^{-\frac32}\,.\notag
% \end{align}
% It follows that
% \begin{align}
% 0&=((-k\cdot\nabla_k)^2+(d+1)k\cdot\nabla_k+m^2\Delta_p+d)\hat{w}(k)\,.\notag
% \end{align}
% The inverse Fourier transform of this gives:
% \begin{align}
% 0&=((\nabla\cdot x)^2-(d+1)(\nabla\cdot x)-m^2|x|^2+d)w(x)\,.\notag
% \end{align}
% Writing $w(x)=|x|^{\frac{1-d}{2}}f(|x|)$ we find
% \begin{align}
% 0&=((d+r\partial_r)^2-(d+1)(d+r\partial_r)-m^2r^2+d)r^{\frac{1-d}{2}}f(r)\notag\\
% &=((r\partial_r)^2+(d-1)r\partial_r-m^2r^2)r^{\frac{1-d}{2}}f(r)\notag\\
% &=r^{\frac{1-d}{2}}((r\partial_r)^2-m^2r^2+\frac{(d-1)^2}{4})f(r)\notag
% \end{align}
% and hence $f(r)$ is a linear combination of the modified Bessel functions $K_{\frac{d-1}{2}}(mr)$ and $I_{\frac{d-1}{2}}(mr)$. Due to the fall-off
% properties at large $r$ we see that $f(r)$ must be a multiple of $K_{\frac{d-1}{2}}(mr)$.

\subsection{Entanglement and separable states}\label{ssec:separablestates}

One of the important properties of the Minkowski vacuum state is that it is highly entangled between any two spacelike separated regions of spacetime \cite{SW1985}. This is a consequence of the Reeh-Schlieder Theorem, which generalises to KMS states, states of bounded energy and many others \cite{RS1961,Haag,Str2000,Wit2018,DM1971,San2009}. Let us briefly review the concept of entanglement in the setting at hand and then consider the problem of constructing states which are separable, i.e.~not entangled.

For any spacetime region $O\subset M$ we denote by $\mathcal{A}(O)$ the subalgebra of $\mathcal{A}$ that is generated by the identity $I$ and elements $\phi(f)$ with $f\in C_0^{\infty}(O)$. When $O_1$ and $O_2$ are two spacelike separated regions (i.e.~for all points $\mathbf{x}\in O_1$ and
$\mathbf{x}'\in O_2$ the vector $\mathbf{x}-\mathbf{x}'$ is spacelike), then we call a state $\omega$ on $\mathcal{A}$ a product state over $\mathcal{A}(O_1)$ and $\mathcal{A}(O_2)$ when
\begin{align}\label{def:productstate}
\omega(A_1A_2)&=\omega(A_1)\omega(A_2)
\end{align}
for all $A_i\in\mathcal{A}(O_i)$. A separable state is a (finite or countably infinite) convex combination of product states, $\omega=\sum_js_j\omega_j$ where $s_j\ge0$, $\sum_js_j=1$ and the $\omega_j$ are product states. A state $\omega$ is entangled between $\mathcal{A}(O_1)$ and $\mathcal{A}(O_2)$ when it is not a separable state.

A quasi-free state $\omega$ with two-point distribution $\omega_2$ is a product state over $\mathcal{A}(O_1)$ and $\mathcal{A}(O_2)$ iff
\begin{align}\label{eqn:qfproductstate}
\omega_2|_{O_1\times O_2}&\equiv 0\,.
\end{align}
This can be seen from Equation (\ref{eqn:qfstate}). Note that $\omega_{2-}$ always vanishes on $O_1\times O_2$ when the $O_i$ are spacelike separated, so when $\omega_2$ vanishes on $O_1\times O_2$ then it also vanishes on $O_2\times O_1$.

One can show from Equation (\ref{eqn:vacuumvalues}) that the Minkowski vacuum state is not a product state between any spacelike separated regions. Moreover, the Minkowski vacuum state violates Bell's inequalities and is therefore entangled \cite{SW1985}. Let us now show, however, that it possible to construct states that are product states between suitable regions of spacetime, using the results of Section \ref{sec:positivity}. Our strategy will be to modify the initial data of the ground state. For convenience we will focus on $d=3$, where we have the following decay properties for the ground state two-point distribution:
\begin{proposition}\label{prop:decay}
For $m>0$ and $d=3$ we write $\omega_{2,00}^{(0)}(x-x')=u(r)$ with $r=|x-x'|$. If $R>0$ we have for all $r\ge R$ and $n\in\Z_{\ge0}$
\begin{align}
\left|\partial_r^nu(r)\right|&\le C_3C_4^{n+3}n!e^{-mr}\,,\label{est:Minkowskifalloff1}
\end{align}
where $C_4=4\max\left\{m,\frac{1}{R}\right\}$ and $C_3=\frac{2}{25m}$.
\end{proposition}
For a proof we refer to Appendix \ref{sec:falloff}.

We now proceed to our main result.
\begin{theorem}\label{thm:main}
Consider a massive real linear scalar quantum field in $3+1$-dimensional Minkowski spacetime and let $R>0$. There exist quasi-free Hadamard states which are homogeneous and isotropic, i.e.~invariant under spatial rotations, and whose two-point distribution vanishes on $\{(\mathbf{x},\mathbf{x}')|\ |x-x'|>R+|x^0-(x')^0|\}$. The renormalised energy density can be made to satisfy
\begin{align}
\omega(E^{\mathrm{ren}}(\mathbf{x}))&\le 10^{31}\frac{m^4}{(mR)^8}e^{-\frac14mR}\,.\notag
\end{align}
\end{theorem}

\begin{proof*}
We fix $R'\in(0,R)$ and let $\chi$ be the characteristic function of the ball $B_{R'}=\{x|\ |x|\le R'\}$ of radius $R'$ in $\R^3$. Next we choose $R''\in(0,R')$ with $R''\le R-R'$ and $\rho>1$. Example \ref{ex:Gevrey2} with $\frac{R''}{\sqrt{3}}$ instead of $R$ then provides a non-negative, rotation invariant test-function $\eta\in C_0^{\infty}(\R^3)$ with support in the ball $B_{R''}$ such that $\int\eta=1$ and
\begin{align}
\|\partial_1^n\eta\|_1&\le\left(\frac{9\rho}{R''(\rho-1)}\right)^nn!^{\rho}\notag
\end{align}
for all $n\in\Z_{\ge0}$. The convolution $\tilde{\chi}:=\eta*\chi$ is smooth, non-negative, rotation invariant, identically $0$ on $|\mathbf{x}|\ge R$ and identically $1$ on $|x|\le R'-R''$. Moreover,
\begin{align}
\|\partial_1^n\tilde{\chi}\|_{\infty}&=\|(\partial_1^n\eta)*\chi\|_{\infty}\le \|(\partial_1^n\eta)\|_1\|\chi\|_{\infty}\le 
\left(\frac{9\rho}{R''(\rho-1)}\right)^nn!^{\rho}\notag
\end{align}
for all $n\in\Z_{\ge0}$ and similarly $\|\partial_1^n(1-\tilde{\chi})\|_{\infty}\le \left(\frac{9\rho}{R''(\rho-1)}\right)^nn!^{\rho}$. Also note that
$1-\tilde{\chi}=1-\eta*\chi=\eta*(1-\chi)$ is identically $1$ on $|x|\ge R$ and identically $0$ on $|x|\le R'-R''$. The spherical symmetry implies
\begin{align}
\|\partial_r^n(1-\tilde{\chi})\|_{\infty}&\le \left(\frac{9\rho}{R''(\rho-1)}\right)^nn!^{\rho}\notag
\end{align}
where $r=|x|$ is the radial coordinate in a spherical coordinate system.

The initial values of the Minkowski vacuum state on the set $x\not=y$ can be written as $\omega^{(0)}_{2,00}(x,y)=u(|x-y|)$ and we see from Proposition \ref{prop:decay} that the smooth function $u(r)$ satisfies
\begin{align}
|\partial_r^nu(r)|&\le C_3C_4^{n+3}n!e^{-mr}\notag
\end{align}
for all $n\in\Z_{\ge0}$ and $r\ge R'-R''$, where $C_3=\frac{2}{25m}$ and $C_4=4\max\{m,\frac{1}{R'-R''}\}$.

Next we define the smooth function $v(x):=(1-\tilde{\chi}(x))\cdot u(|x|)$ which is rotationally invariant and hence $v(x)=\tilde{v}(r)$. We note that for $n\in\Z_{\ge0}$
\begin{align}
|\partial_r^nv(x)|&\le \sum_{j=0}^n\binom{n}{j}|\partial_r^j(1-\tilde{\chi})(x)|\cdot|\partial_r^{n-j}u(r)|\notag\\
&\le C_3C_4^3e^{-mr}\sum_{j=0}^n\frac{n!}{j!}j!^{\rho}\left(\frac{9\rho}{R''(\rho-1)}\right)^jC_4^{n-j}\notag\\
&\le C_5C_6^nn!^{\rho}e^{-mr}\notag
\end{align}
where $C_5:=C_3C_4^3=\frac{128}{25m}\max\{m,\frac{1}{R'-R''}\}^3$,
$C_6:=2\max\{\frac{9\rho}{R''(\rho-1)},C_4\}=8\max\{\frac{9\rho}{4R''(\rho-1)},m,\frac{1}{R'-R''}\}$ and we used the estimate
$\sum_{j=0}^nn!j!^{\rho-1}\le 2^nn!^{\rho}$.

Using the rotational symmetry we have
\begin{align}
\hat{v}(k)
% &=2\pi\int_0^{\infty}\int_0^{\pi} e^{-i|k|r\cos(\theta)}\tilde{v}(r)r^2\sin(\theta)\mathrm{d}\theta\mathrm{d}r\notag\\
% &=2\pi\int_0^{\infty}\left[\frac{r}{i|k|}e^{-i|k|r\cos(\theta)}\tilde{v}(r)\right]_0^{\pi}\mathrm{d}r\notag\\
&=4\pi\int_0^{\infty}\frac{r\tilde{v}(r)}{|k|}\sin(|k|r)\mathrm{d}r\notag
\end{align}
and writing $s_n(r)$ for one of the functions $\pm\sin(r)$ or $\pm\cos(r)$ such that $\partial_r^ns_n(r)=\sin(r)$ we find for all $n\in\mathbb{N}$ that
\begin{align}
|k|^n\hat{v}(k)&=4\pi\int_0^{\infty}\frac{r\tilde{v}(r)}{|k|}\partial_r^ns_n(|k|r)\mathrm{d}r
=4\pi\int_0^{\infty}r\tilde{v}(r)\partial_r^{n-1}s_{n-1}(|k|r)\mathrm{d}r\,.\label{eqn:sn}
\end{align}

Integrating the expression on the right-hand side by parts we find for all $n\ge2$:
\begin{align}
|k|^n|\hat{v}(k)|
% &=||k|^n\hat{v}(k)|\notag\\
&=4\pi\left|\int_0^{\infty}s_{n-1}(|k|r)\partial_r^{n-1}(r\tilde{v}(r))\mathrm{d}r\right|\notag\\
% &\le 4\pi\int_0^{\infty}\left|\partial_r^{n-1}r\tilde{v}(r)\right|\mathrm{d}r\notag\\
&\le 4\pi\int_{R'-R''}^{\infty}\left|r\partial_r^{n-1}\tilde{v}(r)+(n-1)\partial_r^{n-2}\tilde{v}(r)\right|\mathrm{d}r\notag\\
% &\le 4\pi\int_{R'-R''}^{\infty}C_5C_6^{n-1}(n-1)!^{\rho}re^{-mr}+(n-1)C_5C_6^{n-2}(n-2)!^{\rho}e^{-mr}\mathrm{d}r\notag\\
&\le 4\pi C_5C_6^{n-1}(n-1)!^{\rho}\int_{R'-R''}^{\infty}re^{-mr}+C_6^{-1}e^{-mr}\mathrm{d}r\notag\\
&=4\pi C_5C_6^{n-1}(n-1)!^{\rho}\frac{1}{m^2}\left((1+m(R'-R''))+C_6^{-1}m\right)e^{-m(R'-R'')}\notag\\
% &\le 4\pi\frac{17}{16m^2}(1+m(R'-R''))C_5C_6^{n-1}(n-1)!^{\rho}e^{-m(R'-R'')}\,,\label{eqn:estC6b}
&\le C_7C_6^{n-1}(n-1)!^{\rho}e^{-m(R'-R'')}\,,\label{eqn:estC6b}
\end{align}
where $C_7:=\frac{17\pi}{4m^2}C_5(1+m(R'-R''))=\frac{2176\pi}{100}(1+m(R'-R''))\max\{1,\frac{1}{m(R'-R'')}\}^3$ and we used $C_6^{-1}\le\frac{1}{16}\left(\frac{1}{m}+R'-R''\right)$. This estimate also holds for $n=1$, because the second term in the second line vanishes. For $n=0$ we use $|\sin(x)|\le|x|$ to find
\begin{align}
|\hat{v}(k)|&\le 4\pi\int_0^{\infty}\frac{r|\tilde{v}(r)|}{|k|}|\sin(|k|r)|\mathrm{d}r\notag\\
&\le 4\pi\int_0^{\infty}r^2|\tilde{v}(r)|\mathrm{d}r\notag\\
&\le 4\pi C_5\int_{R'-R''}^{\infty}r^2e^{-mr}\mathrm{d}r\notag\\
&\le 4\pi C_5\frac{e^{-m(R'-R'')}}{m^3}(m^2(R'-R'')^2+2m(R'-R'')+2)\notag\\
&\le C_7\frac{2+m(R'-R'')}{m}e^{-m(R'-R'')}\,.\label{eqn:estC6a}
\end{align}

Writing $N(l):=\lfloor\frac{l}{\rho}\rfloor$ and $c=\frac{\rho e^{\frac{1-\rho}{2e}}}{2}$ and distinguishing $\frac{|k|}{C_6}<1$ and $\frac{|k|}{C_6}\ge1$ we find
\begin{align}
e^{cC_6^{-\frac{1}{\rho}}|k|^{\frac{1}{\rho}}}|\hat{v}(k)|&=\sum_{l=0}^{\infty}\frac{c^l}{l!}\left(\frac{|k|}{C_6}\right)^{\frac{l}{\rho}}|\hat{v}(k)|\notag\\
&\le\sum_{l=0}^{\infty}\frac{c^l}{l!}\max\left\{|\hat{v}(k)|,\left(\frac{|k|}{C_6}\right)^{N(l)+1}|\hat{v}(k)|\right\}\notag\\
&\le C_7\frac{e^{-m(R'-R'')}}{m}\sum_{l=0}^{\infty}\frac{c^l}{l!}\max\left\{2+m(R'-R''),C_6^{-1}mN(l)!^{\rho}\right\}\notag\\
&\le C_7(2+m(R'-R''))\frac{e^{-m(R'-R'')}}{m}\sum_{l=0}^{\infty}\frac{c^l}{l!}N(l)!^{\rho}\notag\\
&\le C_7(2+m(R'-R''))\frac{e^{-m(R'-R'')}}{m}2(2\pi)^{\frac{\rho-1}{2}}e^{\frac{13}{12}\rho}\rho^{-\frac{\rho}{2}}\sum_{l=0}^{\infty}\left(\frac12\right)^l\notag\\
&=4(2\pi)^{\frac{\rho-1}{2}}e^{\frac{13}{12}\rho}\rho^{-\frac{\rho}{2}}C_7(2+m(R'-R''))\frac{e^{-m(R'-R'')}}{m}\,,\notag
\end{align}
where we used Equations (\ref{eqn:estC6a},\ref{eqn:estC6b}) in the third line, $\frac{m}{C_6}\le 2+m(R'-R'')$ in the fourth line and we used Stirling's approximation in the form of Lemma \ref{lem:Stirling}. We conclude that
\begin{align}
|\hat{v}(k)|&\le C_8e^{-cC_6^{-\frac{1}{\rho}}|k|^{\frac{1}{\rho}}}\notag
\end{align}
where
\begin{align}
C_8&:=4(2\pi)^{\frac{\rho-1}{2}}e^{\frac{13}{12}\rho}\rho^{-\frac{\rho}{2}}C_7(2+m(R'-R''))\frac{e^{-m(R'-R'')}}{m}\notag\\
&=\frac{2176\pi}{25}(2\pi)^{\frac{\rho-1}{2}}e^{\frac{13}{12}\rho}\rho^{-\frac{\rho}{2}}(1+m(R'-R''))(2+m(R'-R''))
\max\left\{1,\frac{1}{m(R'-R'')}\right\}^3\frac{e^{-m(R'-R'')}}{m}\,.\label{eqn:C8}
\end{align}

Next we will proceed as in Example \ref{ex:Gevrey1} with $\gamma=\frac{1}{\rho}$, $l=\frac{R}{\sqrt{3}}$ and with
$cC_6^{-\frac{1}{\rho}}3^{\frac{1}{2\rho}-1}$ instead of $c$. Thus we let $\{a_n\}_{n\in\N}$ be a positive decreasing sequence such that
$\sum_{n=1}^{\infty}a_n^{\frac{1}{\rho}}$ converges, $a=\sum_{n=1}^{\infty}a_n<\frac{R}{\sqrt{3}}$ and
\begin{align}
\sum_{n=1}^{\infty}a_n^{\frac{1}{\rho}}&\le cC_6^{-\frac{1}{\rho}}3^{\frac{1}{2\rho}-1}(2\rho-1)^{\frac{1}{2\rho}-1}\beta^{-\frac{1}{2\rho}}\notag\\
&=\frac16\rho\left(\frac{15}{56}\right)^{\frac{1}{2\rho}}(2\rho-1)^{\frac{1}{2\rho}-1}e^{\frac{1-\rho}{2e}}
\max\left\{\frac{9\rho}{4R''(\rho-1)},m,\frac{1}{R'-R''}\right\}^{-\frac{1}{\rho}}\,,\notag
\end{align}
where $\beta=\frac{7}{40}$. The test-function $g\in C_0^{\infty}(\R)$ constructed in Proposition \ref{prop:defg} is then positive, even, supported in 
$\left[\frac{-R}{\sqrt{3}},\frac{R}{\sqrt{3}}\right]$ and it satisfies the lower bound
\begin{align}
\hat{g}(k)&\ge e^{-cC_6^{-\frac{1}{\rho}}3^{\frac{1}{2\rho}-1}|k|^{\frac{1}{\rho}}}\,.\notag
\end{align}
As in Section \ref{ssec:higherdbounds} we then define $f_0:=g^{\otimes 3}\in C_0^{\infty}(\R^3)$, which is supported in the ball $B_R$ and satisfies
\begin{align}
\widehat{f_0}(k)&\ge e^{-cC_6^{-\frac{1}{\rho}}|k|^{\frac{1}{\rho}}}\,,\notag
\end{align}
where we used the H\"older Inequality $\displaystyle{\sum_{j=1}^31\cdot|k_j|^{\frac{1}{\rho}}\le 3^{1-\frac{1}{2\rho}}|k|^{\frac{1}{\rho}}}$.
Consequently, the rotation invariant function defined in (\ref{eqn:deff}) also satisfies
\begin{align}
\hat{f}(k)&\ge e^{-cC_6^{-\frac{1}{\rho}}|k|^{\frac{1}{\rho}}}\,.\notag
\end{align}

We now set $w_0(x):=C_8f(x)-v(x)$. For $|x|\ge R$ we note that $f(x)=0$ and $\tilde{\chi}(x)=0$ and hence $w_0(x)=-v(x)=-u(|x|)$. Furthermore, $w_0$ is smooth, rotation invariant and of positive type, because $\widehat{w_0}(k)=C_8\hat{f}(k)-\hat{v}(k)\ge C_8e^{-cC_6^{-\frac{1}{\rho}}|k|^{\frac{1}{\rho}}}-\hat{v}(k)\ge0$. To complete the construction we define $w(x)$ to be the smooth solution to the Klein-Gordon equation with initial data $w(0,x)=w_0(x)$ and
$\partial_0w(0,x)=0$ and we then define the two-point distribution
\begin{align}
\omega_2(x,x')&=\omega^{(0)}_2(x,x')+w(x-x')\,,\notag
\end{align}
which is a bona-fide Hadamard two-point distribution, because $w$ is smooth, real-valued and of positive type, as well as homogeneous and isotropic. The quasi-free state $\omega$ defined by the two-point distribution $\omega_2$ has all the properties stated and it remains to consider its energy density.

From Equation (\ref{eqn:Eren}) we see that the energy density is constant in spacetime and given by
\begin{align}
\omega(E^{\mathrm{ren}}(\mathbf{x}))&\equiv\frac12\left(\sum_{\mu=0}^d\partial_{\mu}\partial_{\mu}'+m^2\right)w(\mathbf{x}-\mathbf{x}')|_{\mathbf{x}'=\mathbf{x}}\notag\\
&=\frac12\left(-\sum_{\mu=0}^d\partial_{\mu}^2+m^2\right)w(\mathbf{x})|_{\mathbf{x}=0}\notag\\
&=\frac12(-\partial_0^2-\Delta+m^2)w(\mathbf{x})|_{\mathbf{x}=0}\notag\\
&=(-\Delta+m^2)w(\mathbf{x})|_{\mathbf{x}=0}\,,\notag
\end{align}
where $\Delta$ is the spatial Laplace operator and we used the Klein-Gordon equation for $w$. The expression on the right-hand side only depends on the initial values on $\Sigma$, which are rotation invariant, so we have
\begin{align}
\omega(E^{\mathrm{ren}}(\mathbf{x}))&\equiv(-\Delta+m^2)w_0(x)|_{x=0}\notag\\
&=C_8(-\Delta+m^2)f(x)|_{x=0}\notag\\
&=C_8(-\Delta+m^2)f_0(x)|_{x=0}\notag\\
&=C_8(-3g''(0)g(0)^2+m^2g(0)^3)\,.\notag
\end{align}
From Proposition \ref{prop:defg} we see that $g(0)\le \frac{3}{2a_1}$ and $-g''(0)\le \frac{9}{a_1^2a_2}$, which entails
\begin{align}
\omega(E^{\mathrm{ren}}(\mathbf{x}))&\le C_8\frac{27}{8a_1^4a_2}\left(18+m^2a_1a_2\right)\,.\notag
\end{align}

Choosing $\rho=2$ and $R'=3R''=\frac34R$ and using (\ref{eqn:C8}) we can express the upper bound as
\begin{align}
\omega(E^{\mathrm{ren}}(\mathbf{x}))&\le m^4\frac{3672\pi}{25}\sqrt{2\pi}e^{\frac{13}{6}}\max\left\{1,\frac{2}{mR}\right\}^3
\left(1+\frac12mR\right)\left(2+\frac12mR\right)e^{-\frac12mR}\frac{18+m^2a_1a_2}{m^5a_1^4a_2}\,.\notag
\end{align}
Choosing $ma_1=ma_2=\frac{1}{523}\min\left\{1,\frac{mR}{18}\right\}$, so that $m(a_1+a_2)<\frac{mR}{\sqrt{3}}$ and
\begin{align}
\sqrt{ma_1}+\sqrt{ma_2}&<\sqrt{m}cC_6^{-\frac{1}{\rho}}3^{\frac{1}{2\rho}-1}(2\rho-1)^{\frac{1}{2\rho}-1}\beta^{-\frac{1}{2\rho}}
=\frac19\left(\frac{45}{56}\right)^{\frac14}e^{\frac{-1}{2e}}\max\left\{\frac{18}{mR},1,\frac{2}{mR}\right\}^{-\frac12}\notag
% Proof:
% $\sqrt{ma_1}+\sqrt{ma_2}=2\sqrt{ma_1}\le 2\frac{1}{\sqrt{523}}\max\left\{1,\frac{18}{mR}\right\}^{-\frac12}$
% and
% $2\frac{1}{\sqrt{523}}\le \frac19\left(\frac{45}{56}\right)^{\frac14}e^{\frac{-1}{2e}}$
% as $LHS<0.0875<RHS$.
\end{align}
For $s=\frac{mR}{2}$ we have on $s>0$ the estimate
\begin{align}
\max\left\{1,\frac{1}{s}\right\}^3(1+s)(2+s)e^{-s}\min\left\{1,\frac{s}{9}\right\}^{-5}\left(18+\frac{1}{523^2}\min\left\{1,\frac{s}{9}\right\}^2\right)
&\le 2\cdot 10^{10}s^{-8}e^{-\frac12s}\notag
\end{align}
% Proof:
% Let $g(s):=s^8\max\left\{1,\frac{1}{s}\right\}^3(1+s)(2+s)e^{-\frac12s}\min\left\{1,\frac{s}{9}\right\}^{-5}\left(18+\frac{1}{523^2}\min\left\{1,\frac{s}{9}\right\}^2\right)$. For $0<s<1$:
% \begin{align}
% g(s)&=9^5(1+s)(2+s)e^{-\frac12s}\left(18+\frac{s^2}{523^2\cdot9^2}\right)\le9^5\cdot 6\cdot 19<10^7\,.\notag
% \end{align}
% For $1\le s<9$:
% \begin{align}
% g(s)&=9^5s^3(1+s)(2+s)e^{-\frac12s}\left(18+\frac{s^2}{523^2\cdot9^2}\right)\le9^5\cdot 6\cdot 19\cdot s^5e^{-\frac12s}\notag\\
% &\le 9^{10}\cdot 6\cdot 19\cdot e^{-\frac92}<10^{10}\,,\notag
% \end{align}
% because the maximum of $s^5e^{-\frac12s}$ on $s\in[1,9]$ occurs at $s=9$. For $s\ge9$ we have
% \begin{align}
% g(s)&=s^8(1+s)(2+s)e^{-\frac12s}\left(18+\frac{1}{523^2}\right)\le\frac{10\cdot 11}{9^2}\left(18+\frac{1}{523^2}\right)s^{10}e^{-\frac12s}\notag\\
% &\le \frac{110}{81}\left(18+\frac{1}{523^2}\right)\left(\frac{20}{e}\right)^{10}<2\cdot 10^{10}\,,\notag
% \end{align}
% because the maximum of $s^{10}e^{-\frac12s}$ on $s\ge9$ occurs at $s=20$.
and therefore
\begin{align}
\omega(E^{\mathrm{ren}}(\mathbf{x}))&\le 10^{31}\frac{m^4}{(mR)^8}e^{-\frac14mR}\,.\notag
\end{align}
\end{proof*}

\begin{remark}
The final estimates in the proof of Theorem \ref{thm:main} may not be sharp. In principle we can minimise the energy density for any given $m,R>0$
w.r.t.~$\rho>1$, $R'\in(0,R)$, $R''\in(0,R')$ and $a_1>a_2>0$ satisfying $R''\le R-R'$, $a_1+a_2<\frac{R}{\sqrt{3}}$ and
$a_1^{\frac{1}{\rho}}+a_2^{\frac{1}{\rho}}<cC_6^{-\frac{1}{\rho}}3^{\frac{1}{2\rho}-1}(2\rho-1)^{\frac{1}{2\rho}-1}\beta^{-\frac{1}{2\rho}}$.
We have refrained from trying to find a (near) minimal estimate, because we have no reason to believe that our construction would get close to the minimal energy density required to have a separable state with the given symmetries. Note, however, that our upper bounds do exhibit some reasonable physical behaviour: the bound has the correct physical units, it diverges polynomially when $mR\to0^+$ and it falls off exponentially when $mR\to\infty$.
% 
% For a field of Planck mass $m=1\sim 10^{-8}$kg we can get a state which is separable over all distances larger than the Planck length
% $R=1\sim 10^{-35}$m if we allow for an energy density (in Planck units) of $10^{31}$, which is $\sim 10^{145}$Jm$^{-3}$.
% 
For a field of proton mass $m=10^{-19}\sim 10^{-27}$kg we can get a state which is separable over all distances larger than a proton radius
$R=10^{20}\sim 10^{-15}$m if we allow for an energy density (in Planck units) of $10^{-54}$, which is $\sim 10^{60}$Jm$^{-3}$.
To get separability over distances longer than $100$ proton radii, however, $R=10^{22}$, we only need to allow for an energy density (in Planck units) of
$10^{-177}$, which is $\sim 10^{-63}$Jm$^{-3}$.
% To get separability over distances longer than $10$ proton radii, $R=10^{21}$, we only need to allow for an energy density (in Planck units) of
% $10^{-71}$, which is $\sim 10^{43}$Jm$^{-3}$.
\end{remark}

Now let us consider separable states in curved spacetimes. Let $M$ be a globally hyperbolic spacetime of dimension four with a trivial topology, with metric $g_{ab}$ and with a given time orientation. By a compact inclusion $\iota:A\to M$ we mean the canonical inclusion of a subset $A\subset M$ such that 
$A$ is open, causally convex (in particular, $A$ is a globally hyperbolic spacetime in its own right) and $\overline{A}\subset M$ is compact with non-empty causal complement. The proof of the following result uses a spacetime deformation argument.
\begin{theorem}\label{thm:curved}
Let $\phi$ be a real scalar field on $M$ of any mass $m\ge0$ and scalar curvature coupling $\xi\in\R$. Let $A, B,C\subset M$ be open regions such that
$A\subset C$, $C$ is spacelike w.r.t. $B$ and the canonical inclusions $\iota_C:C\to M$ and $\iota_A:A\to C$ are compact. Then there exists a quasi-free Hadamard state $\omega$ for $\phi$ which is separable between $A$ and $B$.
\end{theorem}
\begin{proof*}
By Lemma 2.12 of \cite{LS2016} there exists a smooth spacelike Cauchy surface $\Sigma\subset M$ and non-empty open regions $O\subset V\subset\Sigma$ with $\overline{O}\subset V$ and $\overline{V}$ compact, such that $A\subset D(O)$ and $V\subset C$, where $D(O)$ is the domain of dependence of $O$ in $M$. We let $\Sigma_0$ be the $t=0$ Cauchy surface in Minkowski spacetime. Using standard spacetime deformation techniques \cite{FNW1981,Few2015} there now exist the following: (i) globally hyperbolic spacetimes $N_{\pm}$ with Cauchy surfaces $\Sigma_{\pm}$ and isometric, causally convex embeddings $\psi_+:N_+\to M$ and $\psi_-:N_-\to M_0$ such that $\psi_+(\Sigma_+)=\Sigma$ and $\psi_-(\Sigma_-)=\Sigma_0$, (ii) a globally hyperbolic spacetime $\tilde{M}$ and isometric, causally convex embeddings $\tilde{\psi}_{\pm}:N_{\pm}\to\tilde{M}$ such that $\tilde{\Sigma}_{\pm}:=\tilde{\psi}_{\pm}(\Sigma_{\pm})$ are Cauchy surfaces. Furthermore, we can arrange that there exists non-empty open regions $O_0\subset V_0\subset\Sigma_0$ such that $\overline{O_0}\subset V_0$, $\overline{V}_0$ compact and, moreover, $\tilde{\psi}_+(\psi_+^{-1}(O))\subset D(\tilde{\psi}_-(\psi_-^{-1}(O_0)))$ and
$\tilde{\psi}_-(\psi_-^{-1}(V_0))\subset D(\tilde{\psi}_+(\psi_+^{-1}(V)))$ \cite{San2009,Few2015}.

On $M_0$ we now choose a quasi-free Hadamard state $\omega'$ for a real free scalar field of mass $m_0>0$ which is separable between $D(O_0)$ and
$D(\Sigma_0\setminus\overline{V_0})$, using Theorem \ref{thm:main}. On $\tilde{M}$ we consider a free scalar field with a smoothly varying mass $\tilde{m}$ and scalar curvature coupling $\tilde{\xi}$, so that $\tilde{m}=m_0$ and $\tilde{\xi}=0$ on $\tilde{\psi}_-(N_-)$ and $\tilde{m}=m$ and $\tilde{\xi}=\xi$ on
$\tilde{\psi}_+(N_+)$. We use $\tilde{\psi}_-\circ\psi_-^{-1}$ to transport the state $\omega'$ on $M_0$ to a state $\tilde{\omega}$ on $\tilde{M}$, where it suffices to transport the initial data from $\Sigma_0$ to $\Sigma_-$ and then apply the dynamics to define $\tilde{\omega}$. Similarly, we use
$\psi_+\circ\tilde{\psi}_+^{-1}$ to transport the state $\tilde{\omega}$ on $\tilde{M}$ to a state $\omega$ on $M$. Note that $\omega$ is quasi-free and Hadamard (because the deformation is smooth) and by the geometric construction, $\omega_2$ vanishes on $O\times(\Sigma\setminus\overline{V})$. In particular, $\omega$ is separable between $A\subset D(O)$ and $B\subset D(\Sigma\setminus\overline{V})$.
\end{proof*}

Theorem \ref{thm:curved} establishes the existence of separable states between regions $A$ and $B$ under fairly general geometric conditions: $\overline{A}$ and $\overline{B}$ should be spacelike separated and $\overline{A}$ compact. Although the theorem does not provide information about the expected stress tensor of separable states, it does clearly show that separability is not in conflict with the Hadamard property or the quasi-free structure of the state. This indicates that the existence of separable states with these desirable additional properties can be expected as a general feature of quantum fields also in curved spacetimes.

\section{Conclusions}

In spacetimes with trivial topology we have established the existence of separable, quasi-free Hadamard states between spacelike separated regions $A$ and $B$ under reasonable assumptions and for real scalar fields of any mass and scalar curvature coupling. We conjecture that this result can be generalised to spacetimes with non-trivial topology, but additional methods will be required to do so.

For massive fields in Minkowski spacetime we showed that the states can also be stationary, homogeneous and spatially isotropic and separable over all distances $\ge R$ for any chosen $R>0$. In particular, when $R$ is very small, the state looks almost classical. We have shown that the energy density for these states can be bounded by $\le 10^{31}\frac{m^4}{(mR)^8}e^{-\frac14mR}$. We don't expect the constant $10^{31}$ to be sharp and it would be interesting to know by how much this can be decreased. We don't expect that the energy density can be made arbitrarily small. The lowest upper bound in our estimate would be a measure for how much energy it takes to force a quantum field to be separable.

Note that a constant energy density, however small, still leads to an infinite total energy when integrated over all space. A related question, which we did not touch on in this paper, is whether we can also find separable Hadamard states with a finite total energy. This might be expected especially when the regions $A$ and $B$ are bounded, but once again it will require additional methods to settle this question.

\section*{acknowledgements}

I thank Don Page and Stefan Hollands for making me part of an email exchange in which the possibility of creating separable states by cutting down the initial data of the Minkowski vacuum was suggested, without tackling the problem of obtaining distributions of positive type. Initial results were presented at a workshop in Erlangen in 2022 (with weaker estimates) and I thank the organisers and participants for their comments. I am indebted to an anonymous referee for carefully checking the estimates and computations in this paper and in particular for spotting an incorrect constant and a wrong estimate, which have now been corrected.

\appendix

\section{Test functions of Gevrey class}\label{sec:Gevrey}

A test function $f\in C_0^{\infty}(\R^d)$ is said to be in the Gevrey class of order $\rho>1$ if and only if there is a $C>0$ such that
\begin{align}
\sup_{x\in\R^d}|\partial^{\alpha}f(x)|&\le C^{|\alpha|+1}\alpha!^{\rho}\notag
\end{align}
for all multiindices $\alpha=(\alpha_1,\ldots,\alpha_n)$. There are several useful equivalent formulations of this condition:
\begin{proposition}
For $f\in C_0^{\infty}(\R^d)$ and $\rho>1$ the following conditions are equivalent:
\begin{enumerate}[(i)]
\item $f$ is in the Gevrey class of order $\rho$,
\item there is a $C>0$ such that $\displaystyle{\sup_{x\in\R^d}|\partial^{\alpha}f(x)|\le C^{|\alpha|+1}(|\alpha|+1)^{\rho|\alpha|}}$ for all multiindices $\alpha$,
\item there is a $C>0$ such that $\displaystyle{\sup_{k\in\R^d}|k|^N|\hat{f}(k)|\le C^{N+1}(N+1)^{\rho N}}$ for all $N\in\N$,
\item there is a $c>0$ such that $|\hat{f}(k)|\le c^{-1}\exp\left(-c|k|^{\frac{1}{\rho}}\right)$ for all $k\in\R^d$.
\end{enumerate}
\end{proposition}

The proof consists of standard estimates and is omitted, except for the following lemma, which follows from Stirling's approximation and is needed in the main text.
\begin{lemma}\label{lem:Stirling}
For $l\in\N$, $\rho>1$ and $N(l):=\lfloor\frac{l}{\rho}\rfloor$ we have
\begin{align}
\frac{N(l)!^{\rho}}{l!}&\le 2(2\pi)^{\frac{\rho-1}{2}}e^{\frac{13}{12}\rho}\rho^{-\frac{\rho}{2}}\left(\frac{e^{\frac{\rho-1}{2e}}}{\rho}\right)^l\,.\notag
\end{align}
\end{lemma}
\begin{proof*}
For $n\in\N$ we have Stirling's approximation in the form \cite{Robbins}
\begin{align}
\sqrt{2\pi}n^{n+\frac12}e^{-n}e^{\frac{1}{12n+1}}&< n!<\sqrt{2\pi}n^{n+\frac12}e^{-n}e^{\frac{1}{12n}}\,.\notag
\end{align}

When $l\ge\rho$ we have $N(l)\ge1$ and hence
\begin{align}
\frac{N(l)!^{\rho}}{l!}&\le (2\pi)^{\frac{\rho-1}{2}} \left(\frac{l}{\rho}\right)^{l+\frac{\rho}{2}}l^{-l-\frac12} e^{l-\rho N(l)} e^{\frac{\rho}{12N(l)}-\frac{1}{12l+1}}\notag\\
&\le (2\pi)^{\frac{\rho-1}{2}} \rho^{-l-\frac{\rho}{2}} l^{\frac{\rho-1}{2}} e^{\rho+\frac{\rho}{12}}\,.\notag
\end{align}
% The function $g(\rho):=\frac{\rho-1}{2}\log(2\pi)+\frac{13}{12}\rho-\frac{\rho}{2}\log(\rho)$ on $\rho>1$ has a maximum at $\rho=2\pi e^{\frac76}$, where $g(2\pi e^{\frac76})=\pi e^{\frac76}-\frac12\log(2\pi)\le 10$.
% In fact, the value is $\simeq 9.17$.
The function $f(x):=\frac{\rho-1}{2x}\log(x)-\log(\rho)$ on $x\ge1$ has a maximum at $x=e$ where $f(e)=\frac{\rho-1}{2e}-\log(\rho)$.
Therefore,
\begin{align}
\frac{N(l)!^{\rho}}{l!}&\le (2\pi)^{\frac{\rho-1}{2}} e^{\frac{13}{12}\rho}\rho^{-\frac12\rho} e^{lf(l)}
\le (2\pi)^{\frac{\rho-1}{2}} e^{\frac{13}{12}\rho}\rho^{-\frac{\rho}{2}} e^{lf(e)}\notag\\
&=(2\pi)^{\frac{\rho-1}{2}} e^{\frac{13}{12}\rho}\rho^{-\frac{\rho}{2}} \left(\frac{e^{\frac{\rho-1}{2e}}}{\rho}\right)^l\,.\notag
\end{align}
% The extra factor of $2$ in the lemma is not needed in this case.
When $l\in\N$ and $l<\rho$ we distinguish two cases. When $e^{\frac{\rho-1}{2e}}\ge\rho$ we let
$g(\rho):=\frac{\rho-1}{2}\log(2\pi)+\frac{13}{12}\rho-\frac{\rho}{2}\log(\rho)$ with 
$g'(\rho)=\frac12\log(2\pi)+\frac{13}{12}-\frac12-\frac12\log(\rho)$, so $g$ has a maximum at $\rho=2\pi e^{\frac76}$ where
$g(2\pi e^{\frac76})=\frac{-1}{2}\log(2\pi)+\pi e^{\frac76}>0$, so that
\begin{align}
(2\pi)^{\frac{\rho-1}{2}} e^{\frac{13}{12}\rho}\rho^{-\frac{\rho}{2}}&=e^{g(\rho)}\ge 1\,.\notag
\end{align}
Because $N(l)=0$ it then follows that
\begin{align}
\frac{N(l)!^{\rho}}{l!}&=\frac{1}{l!}\le 1\le (2\pi)^{\frac{\rho-1}{2}} e^{\frac{13}{12}\rho}\rho^{-\frac12\rho}\left(\frac{e^{\frac{\rho-1}{2e}}}{\rho}\right)^l\notag
\end{align}
which implies the lemma in this case.
% Again the extra factor of $2$ in the lemma is not needed.

In the case where $e^{\frac{\rho-1}{2e}}<\rho$ we first consider $j(\rho):=\frac{e^{\frac{\rho-1}{2e}}}{\rho}$ on $\rho\ge1$. Note that $j>0$ and
$j'(\rho)=j(\rho)\left(\frac{1}{2e}-\frac{1}{\rho}\right)$, so $j$ has a minimum at $\rho=2e$, where $j(2e)=\frac12e^{-\frac{1}{2e}}$. Furthermore,
$j(\rho)\ge 1$ when $\rho\ge 17$, because $j(17)=\frac{e^{\frac{8}{e}}}{17}>1$ and $j$ is increasing on $\rho>2e$. Now consider the function 
$h(l,\rho):=\frac{\rho}{2}\log(2\pi)+\frac{13}{12}\rho-\frac{\rho}{2}\log(\rho)-l+\frac{2l+1}{2}\log(l)-l\log(2)-\frac{l}{2e}$. We have
$\partial_{\rho}h(l,\rho)=\frac12\log(2\pi)+\frac{13}{12}-\frac12-\frac12\log(\rho)$, which vanishes only at $\rho=2\pi e^{\frac76}>20$, so $h$ is an increasing function of $\rho$ on $\rho\in[1,17]$. It follows that
$h(l,\rho)\ge h(l,1)=\frac12\log(2\pi)+\frac{13}{12}-l+\frac{2l+1}{2}\log(l)-l\log(2)-\frac{l}{2e}$. Now
$\partial_lh(l,1)=\log(l)+\frac{1}{2l}-\log(2)-\frac{1}{2e}$ and
$\partial_l^2h(l,1)=\frac{1}{l}-\frac{1}{2l^2}=\frac{2l-1}{2l^2}>0$ on $l\ge1$. Since
$\partial_lh(2,1)=\log(2)+\frac14-\log(2)-\frac{1}{2e}>0$ we find that for all $l\ge2$, 
$h(l,\rho)\ge h(l,1)\ge h(2,1)=\frac12\log(4\pi)-\frac{11}{12}-\frac{1}{e}>-0.02$. When $l=1$ we have
$h(1,\rho)\ge h(1,1)=\frac12\log(2\pi)+\frac{1}{12}-\log(2)-\frac{1}{2e}>0$. Thus we have for all $l\ge1$, $h(l,\rho)>-0.02$ and 
\begin{align}
\frac{N(l)!^{\rho}}{l!}&=\frac{1}{l!}\le \exp\left(-\frac12\log(2\pi)+l-\frac{2l+1}{2}\log(l)-\frac{1}{12l+1}\right)\notag\\
&\le \exp\left(-h(l,\rho)+\frac{\rho-1}{2}\log(2\pi)+\frac{13}{12}\rho-\frac{\rho}{2}\log(\rho)-l\log(2)-\frac{l}{2e}-\frac{1}{12l+1}\right)\notag\\
&\le e^{0.02}(2\pi)^{\frac{\rho-1}{2}}e^{\frac{13}{12}\rho}\rho^{-\frac{\rho}{2}}\left(\frac12e^{-\frac{1}{2e}}\right)^l\notag\\
&\le e^{0.02}(2\pi)^{\frac{\rho-1}{2}}e^{\frac{13}{12}\rho}\rho^{-\frac{\rho}{2}}\left(\frac{e^{\frac{\rho-1}{2e}}}{\rho}\right)^l\,,\notag
\end{align}
where we noted that $e^{0.02}\le 2$.
\end{proof*}

\section{Decay properties of the Minkowski vacuum state}\label{sec:falloff}

\begin{proof*}[of Proposition \ref{prop:decay}]
The strategy of our proof uses the methods of \cite{FS2014}. For reasons of convenience we will give a complete proof, adapted to our setting and notations.

In $\R^3$ we write $\Delta_r=\frac{1}{r^2}\partial_rr^2\partial_r=\partial_r^2+\frac{2}{r}\partial_r$ for the radial part of the Laplace operator. We note that
\begin{align}
u(r)&=\frac{1}{2(2\pi)^3}(-\Delta_r+m^2)\int e^{irp_1}(m^2+|p|^2)^{-\frac32}\mathrm{d}^3p\notag\\
&=\frac{1}{2(2\pi)^3}(-\Delta_r+m^2)\int e^{irp_1}(m^2+p_1^2)^{-\frac12}\mathrm{d}p_1\int (1+|p'|^2)^{-\frac32}\mathrm{d}^2p'\notag\\
&=\frac{1}{2(2\pi)^2}(-\Delta_r+m^2)\int e^{imrp_1}(1+p_1^2)^{-\frac12}\mathrm{d}p_1\notag\\
&=\frac{1}{(2\pi)^2}(-\Delta_r+m^2)\int_1^{\infty} e^{-mrp}(p^2-1)^{-\frac12} \mathrm{d}p\,,\label{eqn:MinkowskiEvend}
\end{align}
where we exploited the rotation invariance to choose suitable Cartesian coordinates for $p=(p_1,p')$, we rescaled the coordinates $p'$ by $\sqrt{m^2+p_1^2}$ and the final integral uses a contour in the upper half complex plane that goes around the branch cut from $i$ to $\infty$ and that approaches the branch cut from both sides.
% Proof:
% \begin{align}
% \int_0^{\infty} (1+p^2)^{-\frac32}p\mathrm{d}p&=\left[ -(1+p^2)^{-\frac12}\right]_0^{\infty}=1\,.\notag
% \end{align}
% For $a>0$ define $f(z)=e^{iaz}(1+z^2)^{-\frac12}=e^{iaz-\frac12\log(1+z^2)}$, which is holomorphic except when $z=iy$ with $|y|\ge 1$.
% Writing $z=x+iy$ we have $|f(z)|=e^{-ay}|1+z^2|^{-\frac12}$.
% On a half circle with radius $R>1$ we have $|1+z^2|^2\ge (|z|^2-1)^2=(R^2-1)^2$ and hence $|f(z)|\le (R^2-1)^{-\frac12}e^{-ay}$.
% The part of the half circle where $y>\sqrt{R}$ gives a vanishing contribution in the limit $R\to\infty$.
% On the part of the half circle where $y\le\sqrt{R}$ we have $|f(z)|\le (R^2-1)^{-\frac12}$ and the length of this part of the circle is $\sim R^{\frac12}$,
% so this part of the circle also gives a vanishing contribution in the limit $R\to\infty$.
% Near the branch cut with $y>1$, we note that
% \begin{align}
% \lim_{x\to 0^{\pm}}f(x+iy)&=\lim_{x\to 0^{\pm}} e^{ia(x+iy)-\frac12\log(1+(x+iy)^2)}\notag\\
% &=e^{-ay-\frac12\log(y^2-1)-\frac12(\pm i\pi)}\notag\\
% &=\mp ie^{-ay}(y^2-1)^{-\frac12}\,.\notag\\
% \end{align}
% (Note that for $y<1$ the imaginary factor disappears.) Because $f(z)$ has no poles in the upper half plane away from the branch cut
% \begin{align}
% 0&=\int_{-\infty}^{\infty}f(x)\mathrm{d}x - \int_1^{\infty} (-i)e^{-ay}(y^2-1)^{-\frac12} i\mathrm{d}y
% + \int_1^{\infty} ie^{-ay}(y^2-1)^{-\frac12} i\mathrm{d}y\notag\\
% \int_{-\infty}^{\infty}f(x)\mathrm{d}x &= 2\int_1^{\infty} e^{-ay}(y^2-1)^{-\frac12} \mathrm{d}y\,.\notag
% \end{align}

When $r>0$ we can repeatedly differentiate under the integral sign to obtain
\begin{align}
\partial_r^n\int_1^{\infty} e^{-mrp}(p^2-1)^{-\frac12}\mathrm{d}p
&=(-m)^n\int_1^{\infty} e^{-mrp}p^n(p^2-1)^{-\frac12}\mathrm{d}p\notag\\
&=(-m)^ne^{-mr}\int_0^{\infty} e^{-mrp}(p+1)^n(p(p+2))^{-\frac12}\mathrm{d}p\notag
\end{align}
for any $n\in\Z_{\ge0}$.
% Proof:
% For $r>0$ and $0<|h|<\frac12r$ we have $0\le e^{-m(r+h)p}-e^{-mrp}+mphe^{-mrp}\le \frac12m^2p^2h^2e^{-\frac12mrp}$.
% Dividing by $h$, taking absolute values and estimating the integral with a factor $e^{-\frac12mrp}r^l$ for any $l\in\N$ gives
% \begin{align}
% \left|\int_1^{\infty} \frac{e^{-m(r+h)p}-e^{-mrp}+mphe^{-mrp}}{h}p^l(p^2-1)^{-\frac12}\mathrm{d}p\right|
% &\le \frac12|h|m^2\cdot \int_1^{\infty} e^{-\frac12mrp}p^{l+2}(p^2-1)^{-\frac12}\mathrm{d}p\,,\notag
% \end{align}
% where the integral is finite and the result follows when $h\to\0$.
The latter integral can be estimated by splitting the domain into $(0,1)\cup(1,\infty)$ as follows:
\begin{align}
\int_0^1 e^{-mrp}(p+1)^n(p(p+2))^{-\frac12}\mathrm{d}p&\le 2^{n-\frac12}\int_0^1 p^{-\frac12}\mathrm{d}p=2^{n+\frac12}\,,\notag
\end{align}
and
\begin{align}
\int_1^{\infty} e^{-mrp}(p+1)^n(p(p+2))^{-\frac12}\mathrm{d}p&\le\frac{2^n}{\sqrt{3}}\int_1^{\infty} e^{-mrp}p^n\mathrm{d}p\notag\\
&\le\frac{2^n}{\sqrt{3}}(mr)^{-n-1}\int_0^{\infty} e^{-p}p^n\mathrm{d}p
=\frac{2^n}{\sqrt{3}}(mr)^{-n-1}n!\,.\notag
\end{align}
Combining these estimates yields
\begin{align}
\left|\partial_r^n\int_1^{\infty} e^{-mrp}(p^2-1)^{-\frac12}\mathrm{d}p\right|
&\le C_1C_2^nn!e^{-mr}\,,\label{est:Minkowskifalloff2}
\end{align}
for $n\ge1$ and $r\ge R$, where $C_1=\sqrt{2}+\frac{1}{\sqrt{3}mR}$ and $C_2=2\max\{m,\frac{1}{R}\}$.

Using the fact that
\begin{align}
\partial_r^n\frac{1}{r}&=\sum_{l=0}^n\binom{n}{l}(-1)^{n-l}(n-l)!\frac{1}{r^{1+n-l}}\partial_r^l\notag
\end{align}
we find
\begin{align}
\partial_r^nu(r)&=\frac{1}{(2\pi)^2}\partial_r^n\left(-\partial_r^2-\frac{2}{r}\partial_r+m^2\right)\int_1^{\infty}e^{-mrp}(p^2-1)^{-\frac12}\mathrm{d}p\notag\\
&=\frac{1}{(2\pi)^2}\left(-\partial_r^{n+2}-2\sum_{l=0}^n\frac{n!}{l!}(-1)^{n-l}r^{l-n-1}\partial_r^{l+1}+m^2\partial_r^n\right)
\int_1^{\infty}e^{-mrp}(p^2-1)^{-\frac12}\mathrm{d}p\notag
\end{align}
and hence for $r\ge R$, using the estimate (\ref{est:Minkowskifalloff2}),
\begin{align}
|\partial_r^nu(r)|&\le \frac{1}{(2\pi)^2}\left(C_1C_2^{n+2}(n+2)!
+2\sum_{l=0}^n\frac{n!}{l!}R^{l-n-1}C_1C_2^{l+1}(l+1)!+m^2C_1C_2^nn!\right)e^{-mr}\notag\\
&\le \frac{C_1n!}{(2\pi)^2}\left(3C_2^2(2C_2)^n+3C_2^2(2C_2)^n+m^2C_2^n\right)e^{-mr}\notag\\
&\le C_3C_4^{n+3}n!e^{-mr}\,,\notag
\end{align}
where we used $\frac{(n+2)!}{n!}\le 3\cdot2^n$, $\frac{1}{R}\le C_2$ and $\sum_{l=0}^nl+1=\frac{(n+1)(n+2)}{2}\le \frac32\cdot 2^n$ and we set
$C_4:=2C_2=4\max\{m,\frac{1}{R}\}$ and $C_3:=\frac{2}{25m}$ and we note that
\begin{align}
\frac{C_1(6C_2^2+m^2)}{(2\pi)^2}&\le\frac{6\frac14}{(2\pi)^2}C_1C_2^2\le
\frac{25}{4^3\pi^2}\left(\sqrt{2}+\frac{1}{\sqrt{3}}\right)\frac{1}{m}C_4^3\le C_3C_4^3\,.\notag
\end{align}
\end{proof*}

\end{document}